\documentclass[11pt]{article}
\usepackage{graphicx}
\usepackage{amsmath}
\usepackage{amssymb}
\usepackage{epstopdf}

\title{Multiple Schramm-Loewner Evolutions \\ 
  and \\ Statistical Mechanics Martingales} 
\author{}
\date{}

\begin{document}
\maketitle

\vskip -1.0 truecm

\centerline{\large Michel Bauer ${}^{a}$, Denis Bernard ${}^{a}$
\footnote{Member of C.N.R.S},  
Kalle Kyt\"ol\"a ${}^{b}$}
\centerline{\small \texttt{michel.bauer@cea.fr};
\texttt{denis.bernard@cea.fr};
\texttt{kalle.kytola@helsinki.fi}}  

\bigskip

\centerline{${}^a$ \large Service de Physique Th\'eorique de Saclay}
\centerline{CEA/DSM/SPhT, Unit\'e de recherche associ\'ee au 
CNRS} 
\centerline{CEA-Saclay, 91191 Gif-sur-Yvette, France.}
\medskip
\centerline{$^b$ \large Department of Mathematics, P.O. Box 68}
\centerline{FIN-00014 University of Helsinki, Finland.} 

\newcommand{\bC} {\mathbb{C}}
\newcommand{\bR} {\mathbb{R}}
\newcommand{\bZ} {\mathbb{Z}}
\newcommand{\bN} {\mathbb{N}}
\newcommand{\bD} {\mathbb{D}}
\newcommand{\bH} {\mathbb{H}}

\newcommand{\sP} {\mathcal{P}}
\newcommand{\sF} {\mathcal{F}}
\newcommand{\sL} {\mathcal{L}}
\newcommand{\sD} {\mathcal{D}}
\newcommand{\ud} {\mathrm{d}}
\newcommand{\bn} {\mathbf{n}}
\newcommand{\Order} {\mathcal{O}}
\newcommand{\order} {o}
\newcommand{\Oper} {\mathcal{O}}
\newcommand{\unit} {\mathbf{1}}
\newcommand{\Bra} {\Big\langle}
\newcommand{\Ket} {\Big\rangle}
\newcommand{\bra} {\langle}
\newcommand{\ket} {\rangle}
\newcommand{\prob} {\mathbb{P}}
\newcommand{\expect} {\mathbb{E}}
\newcommand{\cconj} {\overline}
\newcommand{\bdry} {\partial}
\newcommand{\im} {\Im \textrm{m }}
\newcommand{\re} {\Re \textrm{e }}
\newcommand{\const} {\mathrm{const.}}
\newcommand{\vir} {\mathfrak{vir}}

\newtheorem{theorem}{Theorem}
\newtheorem{lemma}{Lemma}
\newtheorem{corollary}{Corollary}
\newcommand{\proof} {\emph{Proof: } }
\newcommand{\QED} {$\square$}

\vskip 1.0 truecm

\begin{abstract}
  A statistical mechanics argument relating partition functions to
  martingales is used to get a condition under which random geometric
  processes can describe interfaces in 2d statistical mechanics at
  criticality. Requiring multiple SLEs to satisfy this condition leads
  to some natural processes, which we study in this note. We give
  examples of such multiple SLEs and discuss how a choice of conformal
  block is related to geometric configuration of the interfaces and
  what is the physical meaning of mixed conformal blocks. We
  illustrate the general ideas on concrete computations, with
  applications to percolation and the Ising model.

\end{abstract}

\newpage
\tableofcontents
\newpage

\section{Introduction}

Growth phenomena are ubiquitous around us. They have both very
practical applications and theoretical relevance. But they are rarely
easy to study analytically and very few rigorous or exact results
are known. In two dimensions, the description of a growing domain is
often obtained indirectly through the description of a family of
univalent holomorphic representations, leading quite generally to
equations known under the name of Loewner chains. These techniques,
based on the Riemann mapping theorem, are conceptually important but
usually far from making the problem tractable.

In the last few years, Loewner chains have been discovered which have
a large hidden symmetry -- conformal invariance -- that makes them more
amenable to an exact treatment \cite{schramm0}. These are known
under the name of Stochastic or (Schramm) Loewner evolutions
SLE.

Their mathematical elegance and simplicity is not their sole virtue.
They are also natural candidates to describe the continuum limit of an
interface in two-dimensional statistical mechanics models at
criticality. At the critical temperature and in the continuum limit,
the system is believed to be conformally invariant and physicists have
developed many powerful techniques, known under the name conformal
field theory or CFT, to deal with local questions in a conformally
invariant 2d system. However, nonlocal objects like interfaces posed
new nontrivial problems that finally SLE could attack in a systematic
way \cite{LSW,LSW:ConformalRestriction}. The connection between CFT and
SLE is now well understood
\cite{Bb:2002qn,Bb:2002tf,WernerFrie,Bb:2003kd,Bb:2003vu,Bb:2004ij}
and the interplay between the two approaches has proved fruitful.

The way SLE describes an interface deserves some comments.
As a guiding example, consider the Ising model in a simply connected
domain, say on the hexagonal lattice. Suppose that the boundary is
split in two arcs with endpoints say $a$ and $b$ and impose that on
one arc the spins are up and on the other one the spins are down.  In
this situation each sample exhibits an interface. It joins the two
points where the boundary conditions change and splits the domain
in two pieces, one with all spins up on its boundary and one with all
spins down. This interface fluctuates from sample to sample. What
SLE teaches us is the following. Instead of describing the
interface between $a$ and $b$ at once, SLE views it as a curve
starting from say $a$ and growing toward $b$. And SLE describes the
distribution for the addition of an infinitesimal piece of interface
when the beginning of the interface is already known.  So the
description is in terms of a growth process even if there was no
growth process to start with.

As mentioned above, the probabilistic aspects of SLE as well
as its connections with conformal field theory are now fairly well
understood.  However, some fundamental questions remain, again
directly related to natural questions in the statistical mechanics
framework.

The one we shall concentrate on in this note is what happens when, due
to boundary conditions, the system contains several interfaces.
Proposals for multiple SLEs have already been made in the literature
\cite{Cardy:nSLE,D:SLEcommutation}, but our results point to a
different picture. The simplest situation is in fact when there is
only one interface but we want to deal with its two ends symmetrically
so that two growth processes will interact with each other. Remember
that standard SLE deals with the two ends of the interface
asymmetrically. This has a price : time reversed SLE is an intricate
object.

As a guiding example for more than one interface, consider again the
Ising model in a simply connected domain on the hexagonal lattice. If
one changes boundary conditions from up to down to up and so on $n=2m$
times along the boundary, each sample will exhibit $m$ interfaces,
starting and ending on the boundary at points where the boundary
conditions change, forming a so-called arch system. However, the
interfaces will fluctuate from sample to sample and so does even the
topology of the arch system. This topology, for instance, is an
observable that is trivial for a single SLE.

Our description will again be in terms of growth processes and Loewner
chains. For standard SLE, the driving parameter is a continuous
martingale and the tip of the curve separates two different states of
the system (up and down spins for Ising), leading to a well defined
boundary changing operator in statistical mechanics. The relation
between the stochastic Loewner equation and the boundary changing
operator comes via a diffusion equation that they share in common.

For multiple SLEs, we expect that for short time scales each curve
grows under the influence of an independent martingale. At its tip
stands the same boundary changing operator. But we also expect drift
terms, describing interactions between the curves.

The possibility of different arch topologies makes it even more
natural to have a description with one curve growing at each boundary
changing point so that each of them is on the same footing. So $m$
interfaces are described by $n=2m$ growth processes of ``half-interfaces''
that finally pair in a consistent way to build arches. 

In statistical mechanics, each arch system has a well defined
probability to show up. The law governing this finite probability
space is again described by a Boltzmann weight which is nothing but a
partial partition function.

Our starting point is the reconsideration of the role of Boltzmann
weights and partition functions in statistical mechanics and their
simple but crucial relationship with probabilistic martingales.  This
allows us to ask the question ``by what kind of stochastic
differential equations can one describe multiple SLEs ?'' by imposing
a martingale property and conformal invariance. This puts strong
constraints on the drift terms and our main result is a description
of the family of drift terms that are compatible with the basic rules
of statistical mechanics. Each drift is expressed in terms of the
partition function of the system. This partition function is given by
a sum of Boltzmann weights for configurations that satisfy certain
boundary conditions : at the starting points of the curves the
boundary conditions change.  The partition function depends on the
position of these changes, so up to normalization, the partition
function is in fact a correlation function. It satisfies a number of
partial differential equations (one equation for each point) that are
related to the diffusion equations for the multi SLE process. The
solutions form a finite dimensional vector space. The positivity
constraint satisfied by physical partition functions singles out a
cone which is expected, again guided by statistical mechanics, to have
the same dimension of the underlying vector space and to be the convex
hull of a family of half 
lines, so that a generic hyperplane section of the cone is a simplex. 

So geometrically, the drift terms are parametrized by a cone.
Extremal drifts, i.e. drifts corresponding to extremal lines in this
cone, lead to processes for which the final pattern formed by the
growing curves is a given arch system. Drifts inside the simplex give
rise to stochastic processes where the asymptotic arch system
fluctuates from sample to sample. A crucial role to construct
martingales describing interesting events is played by the short
distance expansion in conformal field theory because this is what
tells which terms in the partition function become dominant when an
arch closes, i.e. when two driving processes of the multi SLE hit each
other. 

The vector space of solution of the differential equations
for the partition functions has a famous basis indexed by Dyck
paths, which are in one to one correspondence with arch systems. But
the basis elements do not in general correspond to extremal partition
functions. We shall give a  rationale for computing the matrix
elements for the change of basis and compute a number of them, but we
have no closed general formula.

We shall illustrate our proposal with concrete computations for $1$ to
$3$ interfaces with applications to percolation and the Ising model. We
shall also discuss the classical (deterministic) limit $\kappa
\rightarrow 0^+$, where only extremal drifts survive.

The notes also cover the case when a number of boundary changes are
very close to each other but the system is conditioned so that they
do not pair with each other. The details are in the main text.

Our description is rather flexible in the sense that the speed of
growth of each piece of interface can be tuned. Certain limiting cases
lead to previously known processes which are examples of
SLE$(\kappa,\underline{\rho})$ processes.

It is appropriate here to stress that many of the probabilistic
properties of the solutions of the stochastic differential equations
that we introduce are conjectural at this point. We have made some
consistency checks\footnote{For instance, Dub\'edat has derived
  general ``commutation criteria'' \cite{D:SLEcommutation} for
  multiple SLEs.  The processes we study are a special class
  satisfying commutation.  This class extends vastly the special
  solution found by Dub\'edat, which in our language corresponds to
  self avoiding SLEs moving to infinity.} and the whole pattern is
elegant, but our confidence comes more from our familiarity with
conformal field theory and statistical mechanics.

\section{Basics of Schramm-L\"owner evolutions: Chordal SLE}

Let us briefly recall what is meant by the chordal SLE --- detailed
studies can be found in \cite{RS:BasicProperties} or \cite{W:Lectures}.
The chordal SLE process in the upper half plane $\bH$
is defined by the ordinary differential equation
\begin{equation}
\label{eq: chordal SLE}
\frac{\ud}{\ud t} g_t(z) = \frac{2}{g_t(z)-\xi_t}
\end{equation}
where the initial condition is $g_0(z) = z \in \bH$ and
$\xi_t = \sqrt{\kappa} B_t$ is a Brownian motion with variance
parameter $\kappa \geq 0$. Let $\tau_z \leq \infty$ denote the
explosion time of (\ref{eq: chordal SLE}) with initial condition
$z$ and define the hull at
time $t$ by $K_t := \overline{ \{ z \in \bH | \tau_z < t  \} }$.
Then $(K_t)_{t \geq 0}$ is a family of growing hulls,
$K_s \subset K_t$ for $s<t$.
The complement $\bH \setminus K_t$ is
simply connected and $g_t$ is the unique conformal mapping
$\bH \setminus K_t \rightarrow \bH$ with
$g_t(z) = z + \order (1)$ at $z \rightarrow \infty$. One
defines the SLE trace by $\gamma_t = \lim_{\epsilon \downarrow 0}
g_t^{-1}(\xi_t + i \epsilon)$. The trace is a continuous
path in $\overline{\bH}$ and it generates the hulls in the sense
that $\bH \setminus K_t$ is the unbounded component of
$\bH \setminus \gamma_{[0,t]}$. For $\kappa \leq 4$ the trace
is a non-self-intersecting path and it doesn't hit
$\partial \bH = \bR$ for $t>0$ so $K_t = \gamma_{[0,t]}$. For
$4 < \kappa < 8$ a typical point $z \in \bH$ is swallowed, i.e.
$z \in K_t$ for large $t$ but $z \notin \gamma_{[0, \infty)}$.
In the parameter range $\kappa \geq 8$ the trace is space
filling, $\gamma_{[0, \infty)} = \overline{\bH}$. Let us point
out that no statistical mechanics models seem to correspond
to $\kappa > 8$.

In the definition of chordal SLE we took the usual parametrization of
time. From equation (\ref{eq: chordal SLE}) we see that $g_t (z) = z +
2 t z^{-1} + \Order(z^{-2})$, which means (this could be taken as a
definition) that the capacity of $K_t$ from infinity is $2 t$.  Since
the capacity goes to infinity as $t \rightarrow \infty$, the hulls
$K_t$ are not contained in any bounded subset of $\overline{\bH}$. 

If the parametrization of time is left arbitrary, Schramm's argument
yields :
\[ \ud g_s(z) = \frac{2\ud q_s}{g_s(z)-M_s},\]
where $M_s$ is a continuous martingale with quadratic variation
$\kappa q_s$ (an increasing function going to infinity with $s$). In
this formula, both $q_s$ and $M_s$ are random objects. The capacity of
$K_s$ is $2q_s$. But this is not really more general than eq.(\ref{eq:
  chordal SLE}) which is recovered by a random time change.

\section{A proposal for multiple SLEs}

The motivations for our proposal require a good amount of background,
but the proposal and its main features themselves can be easily
stated. We gather them in this section. Some of the results are
conjectures. The rest of the paper will then be split into sections
whose purpose will be either to motivate our proposal in general, or
to prove its correctness in certain special but nontrivial cases by
explicit computations.

\subsection{The basic equations}

We propose to describe the local growth of $n$ interfaces in CFT,
labeled by an integer $i=1,\cdots,n$ and joining fixed points on the
boundary by a Loewner chain. We assume that $0 \leq \kappa <8$ in the
following. We list the set of necessary conditions and equations.

\vspace{.3cm}

\noindent \textbf{Conformal invariance}: The measure on $n$SLE is
conformally invariant. Hence it is enough to give its definition when
the domain $D$ is the upper half plane $\mathbb H$ in the
hydrodynamical normalization.

\vspace{.3cm}

\noindent \textbf{Universe}: The basic probabilistic objects
are $n$ (continuous, local) martingales $M^{(i)}_t$, $i=1,\cdots,n$
with quadratic variation $\kappa q^{(i)}_t$ absolutely continuous with
respect to $\ud t$ and vanishing cross variation, defined on an
appropriate probability space. By a time change we can and shall
assume that $\sum_i q^{(i)}_t \equiv t$.

\vspace{.2cm}

\noindent \textbf{Driving processes}: The  processes $X^{(i)}_t$ are
solutions of the stochastic differential equations
\begin{equation}
\label{eq:drivepro}
\ud X^{(i)}_t=\ud M^{(i)}_t + \kappa \ud
q^{(i)}_t(\partial_{x_i}\log Z) (X^{(1)}_t,\cdots,X^{(n)}_t)+ \sum_{j
  \neq i} \frac{2\ud q^{(j)}_t}{X^{(i)}_t-X^{(j)}_t}.
\end{equation}
The initial
conditions are $X_0^{(i)}=X_i$ ordered in such a way that $X_1 <
X_2< \cdots < X_n$.
 
\vspace{.2cm}

\noindent \textbf{Loewner chain}: The map $f_t$ uniformizing the
complement of the hulls satisfies
\begin{equation}
\label{eq:loewchain}
\ud f_t(z)=\sum_i \frac{2\ud
    q^{(i)}_t}{f_t(z)-X^{(i)}_t}.
\end{equation}
The initial condition is $f_0(z)=z$. With our conventions, the total
capacity of the growing hulls at time $t$ is $2t$.

\vspace{.2cm}

\noindent \textbf{Auxiliary function}: The system depends on a function
$Z(x_1,\cdots,x_n)$ which has to fulfill the following requirements :

$i)$ $Z(x_1,\cdots,x_n)$ is defined and positive for $x_1 < x_2< \cdots
< x_n$,

$ii)$ $Z(x_1,\cdots,x_n)$ is translation invariant and homogeneous.
Its weight is $h_{n-2m}(\kappa)-nh_{1}(\kappa)$ for some
nonnegative integer $m \leq n/2$, where\footnote{A more traditional
  notation for $h_{m}(\kappa)$ is $h_{1,m+1}$ in the physics literature.} 
$$2\kappa h_{m}(\kappa)\equiv
m(2(m+2)-\kappa).$$

$iii)$ $Z(x_1,\cdots,x_n)$ is annihilated by the $n$ differential
operators
$${\mathcal D}_i=\frac{\kappa}{2}\partial_{x_i}^2+2\sum_{j \neq i}\left[
\frac{1}{x_j-x_i}\partial_{x_j}-\frac{h_{1}(\kappa)}{(x_j-x_i)^2}\right].$$

\vspace{.2cm}

We call this system of equations the $n$SLE system for $n$ curves
joining together the points $X_1,\cdots,X_n$ and possibly the point at
infinity. Systems for radial and dipolar versions of $n$SLE
could be defined analogously.

\subsection{Arch probabilities}

It is known from CFT that (relaxing the positivity constraint), the
solutions to $i),\;ii),\;iii)$ form a vector space of dimension
$d_{n,m}\equiv {n \choose m}-{n \choose
  m-1}=(n+1-2m)\frac{n!}{m!(n-m+1)!}$.

The positive solutions form a cone and from the statistical mechanics
interpretation, we conjecture that this cone has the same dimension
and is generated by (i.e. is the convex hull of) $d_{n,m}$ half 
lines (extremal lines, pure states in the sense of
statistical mechanics) so that a transverse section of the cone is a
simplex. So each solution $Z$ can be written in a unique way as a sum of
extremal states.

The numbers $d_{n,m}$ have many many combinatorial interpretations,
but the one relevant for us is the following. Draw $n+1$ points
$X_1<X_2\cdots <X_n < \infty$ ordered cyclically on the (extended)
real line bounding the upper half plane $\mathbb H$. Consider $n-m$
disjoint curves in $\mathbb H$ such that each $X_i$ is an end
point of exactly $1$ curve and $\infty$ is an end point of exactly
$n-2m$ curves. There are $d_{n,m}$ topologically inequivalent
configurations, called arch configurations when $n-2m=0$. We keep the
same name for $m\neq 0$, writing arch$_m$ configurations when
precision is needed.

Motivated by this, we claim the following :

\vspace{.3cm}

a) To each arch configuration $\alpha$ corresponds an
extremal state $Z_{\alpha}$ in the following sense : the solution of the
$n$SLE system with $Z \propto Z_{\alpha}$ can be defined up to a
(possibly infinite) time, at which the growing curves have either paired
together or joined the point at infinity and at that time the topology is
that of the arch $\alpha$ with probability one.

\vspace{.2cm}
 
b) One can decompose a general solution $Z$ of $i),\;ii),\;iii)$ as a sum
of  $$\sum_{\alpha \,\in \, {\mathrm{arch}_m}} Z_{\alpha}.$$

\vspace{.2cm}

c) The probability that a solution of the $n$SLE system with auxiliary
function $Z$ ends in arch configuration $\alpha$ is the ratio
$$\frac{Z_{\alpha}(X_1,\cdots,X_n)}{Z(X_1,\cdots,X_n)}$$
evaluated at the initial condition $(X_1,\cdots,X_n)$.

\vspace{.3cm}

The first step toward a heuristic derivation of the above results will
be to explain how to construct martingales -- in particular
martingales associated to interfaces -- from statistical mechanics
observables in a systematic way. But we start with a few comments.

\section{First comments}

\subsection{Statistical mechanics interpretation}

To have a specific example in mind, think again of the Ising model at
the critical temperature. Let $a$ be the lattice spacing.  

First, put $n=2m$ changes of boundary conditions from spins up to
spins down and so on along the boundary at points
$x_1/a,\cdots,x_n/a$. In the continuum limit when $a \rightarrow 0$
but $x_1,\cdots,x_n$ have a finite limit, the partition function
behaves like a homogeneous function $Z(x_1/a,\cdots,x_n/a)$ of weight
$0$ (when both $a$ and the $x_i$'s are rescaled) and CFT teaches us
that $Z(x_1,\cdots,x_n)$ satisfies $i),\;ii),\;iii)$ for $n=2m$. Then,
if $\alpha$ is an arch system, $Z_{\alpha}$ should be (proportional to
the continuum limit of) the partial partition function when the sum of
Boltzmann weights is performed only over the interface configurations
with topology $\alpha$.

To make generalized arch configurations, choose $n$ and $m$ with
$n\geq 2m$. Put $2n-2m$ changes of boundary conditions from spins up
to spins down and so on along the boundary, $n$ at points
$x_1/a,\cdots,x_n/a$ and $n-2m$ at $y_1/a,\cdots, y_{n-2m}/a$. Sum
only over configurations where the interfaces do not joint two
$y$-type points to each other. Take the continuum limit for the $x$'s
as before, but impose that all $y$'s go to infinity and remain at a
finite number of lattice spacings from each other. This is expected to
lead again to a partition function $Z(x_1/a,\cdots,x_n/a)$ of weight
$0$ (when both $a$ and the $x_i$'s are rescaled) and
$Z(x_1,\cdots,x_n)$ satisfies $i),\;ii),\;iii)$ for the given $n$ and
$m$. If $\alpha$ is an arch$_m$ configuration, $Z_{\alpha}$ should be
(proportional to the continuum limit of) the partial partition
function when the sum of Boltzmann weights is performed only over the
interface configurations with topology $\alpha$.

Note that the prefactor between the continuum limit finite part and
the real partition function is a power of the lattice spacing. The
power depends on $m$, so it is likely to be unphysical to use a non
homogeneous $Z$ in the $n$SLE system, mixing different values of $m$
for a fixed $n$. However, we shall later treat the example $n=2$
mixing $m=0$ and $m=1$ because it is illustrative despite the fact
that it breaks scale invariance. 

\subsection{SLE as a special case of $2$SLE}

For $n=2$ the solution of $i),\;ii),\;iii)$ with $m=1$ is elementary.
Writing $x_1=a$ and $x_2=b$, one finds $Z\propto (b-a)^{(\kappa
  -6)/\kappa}$. Taking the first martingale to be a Brownian and the
second one to be $0$, one retrieves the equations for SLE growing from
point $a$ to point $b$ in the hydrodynamical normalization.  Let us recall
briefly why.
 
We start from SLE from $0$ to $\infty$.  The
basic principle of conformal invariance makes the passage from this
special case to the case when SLE goes from point $a$ to point $b$ on
$\bH$ a routine task. If $u$ is any linear fractional transformation
(i.e. any conformal transformation) from $\bH$ to itself mapping $0$
to $a$ and $\infty$ to $b$, the image of the SLE trace or hull from
$0$ to $\infty$ by $u$ is by definition an SLE trace from $a$ to $b$
and this is measure preserving. The new uniformizing map is
$h_t=u\circ g_t \circ u^{-1}$ and it is readily checked that
$\frac{\ud h_t}{\ud t}$ is a rational function of $h_t$ whose precise
form can be easily computed but does not concern us.

Let us just mention that this rational function is regular everywhere
(infinity included) except for a simple pole at $h_t=u(\xi_t)$ and
has a third order zero at $h_t=u(\infty)=b$. So the map $h_t$ is
normalized in such a way that
$h_t(b+\varepsilon)=b+\varepsilon+O(\varepsilon^3)$, which is not the
hydrodynamic normalization.

But if $v_t$ is any linear fractional transformation, $v_t \circ g_t
\circ u^{-1}$ describes the same trace as $h_t=u \circ g_t \circ
u^{-1}$. As long as the trace does not separate $b$ from $\infty$,
i.e. as long as the trace has not hit the real axis in the segment
$]b,\infty[$, i.e. as long as $\infty$ is not in the hull, $v_t$ can
be adjusted in such a way that $\tilde{h}_t\equiv v_t \circ g_t \circ
u^{-1}$ is normalized hydrodynamically. Then $\ud \tilde{h}_t/\ud t$
is a function of $\tilde{h}_t$ which is regular everywhere but for a
single pole and vanishes at infinity, i.e. one can write $\ud
\tilde{h}_t/\ud t=2\mu_t/(\tilde{h}_t-\alpha_t)$.  The following
computation is typical of the manipulations made with SLE (see e.g.
\cite{LSW:ConformalRestriction}).  Write $(g_t \circ u^{-1})(z)=w$ an
compute from the definition
$$\frac{\ud \tilde{h}_t}{\ud t}(z)=\frac{\ud
  v_t}{\ud t}(w)+ v_t^{'}(w)\frac{2}{w-\xi_t}.$$ Comparison gives
$$\frac{\ud v_t}{\ud
  t}(w)=\frac{2\mu_t}{v_t(w)-\alpha_t}-\frac{2v_t^{'}(w)}{w-\xi_t}.$$
But $v_t$ is regular at $w=\xi_t$ from which one infers that
$v_t(\xi_t)=\alpha_t$ (the poles in the two terms are at the same
point) and $\mu_t=v_t^{'}(\alpha_t)^2$ (the two residues add to $0$).
Going one step further in the expansion close to $\xi_t$ yields
$\frac{\ud v_t}{\ud t}(\xi_t)= -3v_t^{''}(\xi_t).$ Ito's formula gives
$\ud \alpha_t=-3v_t^{''}(\xi_t)\ud t+v_t^{'}(\xi_t)\ud
\xi_t+\frac{\kappa}{2}v_t^{''}(\xi_t)\ud t.$ So the time change $\mu_t
\ud t =\ud s $ together with the definition $\ud \chi_s
=v_t^{'}(\xi_t)\ud \xi_t$ yields
$$\ud \alpha_t(s)=\ud
\chi_s+(\kappa-6)\frac{v_t^{''}(\xi_t)}{2v_t^{'}(\xi_t)^2}\ud s.$$
But
$v_t^{''}(w)/v_t^{'}(w)^2=2/(v_t(w)-v_t(\infty))$ because $v_t$ is a
linear fractional transformation.  Finally, setting
$\tilde{h}_{t(s)}\equiv f_s$,
$\tilde{h}_{t(s)}(b)=v_{t(s)}(\infty)\equiv B_s$ and
$v_{t(s)}(\xi_{t(s)})=\alpha_{t(s)}\equiv A_s$ we can summarize
$$\frac{\ud f_s}{\ud s}=\frac{2}{f_s-A_s}\; , \quad \frac{\ud B_s}{\ud
  s}=\frac{2}{B_s-A_s} \; , \quad \ud A_s= \ud \chi_s
+(\kappa-6)\frac{\ud s}{A_s-B_s},$$
where $\chi_s$ is a Brownian
motion with quadratic variation $\kappa s$, $f_0=id$, $A_0=a$,
$B_0=b$. Thus chordal SLE from $a$ to $b$ in the hydrodynamical
normalization is indeed a special case of $2$SLE.

The above equations are also a special case of SLE$(\kappa,\rho)$
$(\rho=\kappa-6)$, but it should be clear that our general
proposal goes in a different direction.

As already mentioned, the description of chordal SLE from $a$ to $b$
in the hydrodynamical normalization in fact coincides with chordal SLE
from $a$ to $b$ only up to the first time $b$ is separated from
$\infty$ by the trace. This time is infinite for $\kappa \leq 4$, but
it is finite with probability $1$ for $4 < \kappa < 8$. The most
obvious case is $\kappa=6$. The equation is nothing but the usual
chordal SLE$_6$ ending at infinity, a consequence of locality (in the
SLE sense, not in the quantum field theory sense used later). At that
time, the real chordal SLE from $a$ to $b$ swallows $\infty$, whereas
the hydrodynamically normalized version swallows $b$. The solution to
this problem is to use conformal invariance and restart the process
again in the correct domain at the time when $b$ and $\infty$ get
separated by the trace. But this is not coded in the equations.

\subsection{Making sense}

The previous example should serve as a warning. Some serious
mathematical work may have to be done even to make sense of our
conjectures, let alone prove their correctness. The problems might be
of different natures for $\kappa \leq 4$ and $4 < \kappa <8$.  We
content with the following naive remarks. One of the problems is that
the arches do not have to close at the same time. It may even happen
that one of the growing curves touches the real line or another curve
in such a way that the upper half plane is split in two domains and
the one which is swallowed contains some of the growing curves.

\begin{figure}
\includegraphics[width=1.0\textwidth]{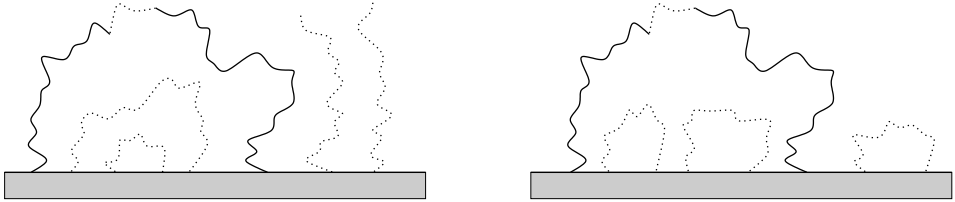}
\caption{\emph{The probability of closing of an arch should be obtained
    by summing $Z_\alpha/Z$ over all $\alpha$'s corresponding to
    compatible configurations. Two compatible configurations are
    portrayed in the figure.}}
\label{fig: compatible configurations}
\end{figure}

Our putative description of $n$SLE processes can be valid in this form
only up the realization of such an event. The first thing to check
should be that the event is realized with a probability obtained by
summing $Z_{\alpha}/Z$ over all $\alpha$'s corresponding to compatible
configurations (see figure \ref{fig: compatible configurations}).
In particular, the connected
component of $\infty$ should contain at least $n_{\infty} \geq m-1$
curves for consistency, but that's not an obvious property of our
proposal.

Consider the fate of the connected component of $\infty$. If
$n_1-m$ is even, conformal invariance suggests to continue the Loewner
evolution simply by suppressing the points that have been swallowed,
i.e. for the $n_1$ remaining points. If $n_1-m$ is odd, the same
should be done, but the image of the point at which one interface has
made a bridge should be included as a starting point for the
continuation of the evolution.  Preferably, the function $Z$ for this new
$multi$SLE system should not be adjusted by hand to make our
conjectures correct, but should appear as a natural limit.  We shall
make comments on this and give concrete illustrations later.

For the component that is
swallowed, one can use conformal invariance again to change the
normalization of the Loewner map in such a way that this component is
the one that survives and then restart a new $multi$SLE for the
appropriate number of points. This procedure may have to be iterated.

Note also that our conjectures for arch probabilities do not involve
any details on the martingales $M^{(i)}_t$. Indeed, we expect that
there is some robustness. But the precise criteria are beyond our
understanding.

\subsection{A few martingales for $n$SLEs} 

Our heuristic derivation of the $n$SLE system will in particular show
that if $\tilde{Z}$ also solves $i),\;ii),\;iii)$ (even relaxing
positivity), the quotient
$$\frac{\tilde{Z}(X^{(1)}_t,\cdots,X^{(n)}_t)}
{Z(X^{(1)}_t,\cdots,X^{(n)}_t)}$$ 
is a local martingale. This can be proved directly using Ito's formula. 

In particular,
$$\frac{Z_{\alpha}(X^{(1)}_t,\cdots,X^{(n)}_t)}
{Z(X^{(1)}_t,\cdots,X^{(n)}_t)}$$
is a local martingale bounded by
$1$, hence a martingale. On the other hand, a
standard argument shows that if $P_{\alpha}$ is the probability that
the system ends in a definite arch configuration $\alpha$ (once one
has been able to make sense of it)
$P_{\alpha}(X^{(1)}_t,\cdots,X^{(n)}_t)$ is a martingale.  This is an
encouraging sign. To get a full proof, one would need to analyze the
behavior of $Z_{\alpha}(X^{(1)}_t,\cdots,X^{(n)}_t)$ when one arch
closes, or when one growing curve cuts the system in two, to get
recursively a formula that looks heuristically like
$$\frac{Z_{\alpha}(X^{(1)}_t,\cdots,X^{(n)}_t)}
{Z(X^{(1)}_t,\cdots,X^{(n)}_t)}\sim \delta_{\alpha,\alpha'}$$
if the
system forms asymptotically the arch system $\alpha'$ at large $s$.
Such a formula rests on properties of $Z_{\alpha}(x_1,\cdots,x_n)$
when some points come close together in a way reminiscent to the
formation of arch $\alpha'$ : $Z_{\alpha'}(x_1,\cdots,x_n)$ should
dominate all $Z_{\alpha}$'s, $\alpha \neq \alpha'$ in such
circumstances. In section \ref{sec:sevinterf} we shall use this to
expand explicitly the $Z_{\alpha}$'s in a basis of solutions to
$i),\;ii),\;iii)$ which is familiar from CFT, very explicitly at least
for small $n$.

\subsection{Classical limit} 

Our proposal for $n$SLE has a non trivial classical limit at $\kappa
\rightarrow 0^+$. The martingales $M_t^{(i)}$ vanish in this limit,
but the $q_t^{(i)}$ remain arbitrary increasing functions. The
function $Z$ does not have a limit, but the $U_i\equiv \kappa
\partial_{x_i} \log Z$ do. They are kind of Ricatti variables for
which the equations read
$$
\frac{1}{2} \left(\partial_{x_i}U_i +
  \frac{U_i^2}{\kappa}\right)+2\sum_{j \neq i}
\left(\frac{1}{x_j-x_i}\frac{U_i}{\kappa}-
  \frac{6-\kappa}{2\kappa}\frac{1}{(x_j-x_i)^2}\right)=0,
$$ 
which have a limit when $\kappa \rightarrow 0^+$, comparable to the
classical limit of a Schroedinger equation. To summarize, the
classical limit is 

$$\ud f_t(z)=\sum_i \frac{2\ud q^{(i)}_t}{f_t(z)-X^{(i)}_t}.$$
$$
\ud X^{(i)}_t=U_i(X^{(1)}_t,\cdots,X^{(n)}_t) \ud q^{(i)}_t
    +\sum_{j \neq i} \frac{2\ud q^{(j)}_t}{X^{(i)}_t-X^{(j)}_t}.$$
where the auxiliary
functions $U_i(x_1,\cdots,x_n)$ are homogeneous functions of degree
$-1$ which satisfy $\partial_{x_i}U_j=\partial_{x_j}U_i$ and
$$ \frac{1}{2} U_i^2+2\sum_{j \neq i}\left(\frac{1}{x_j-x_i}U_i-
  \frac{3}{(x_j-x_i)^2}\right)=0.$$

It is not too surprising that the differential equations for $Z$ have
become algebraic equations for the $U_i$'s, so that the space of
solutions which was a connected manifold for $\kappa \neq 0$
concentrates on a finite number of points in the classical limit.  The
classical system, maybe with an educated guess for the $q_t^{(i)}$'s,
could be interesting for its own sake.

\subsection{Relations with other work}

Several processes involving several growing curves have appeared
in the literature.  

The first proposal was made by Cardy \cite{Cardy:nSLE}. It can be formally
obtained from ours by forgetting the conditions $i),\;ii),\;iii)$ and
choosing a constant $Z$. The corresponding processes are interesting,
but the relationship with interfaces in statistical mechanics and CFT
is unclear for us.

Dub\'edat \cite{D:SLEcommutation} has derived a general criterion he calls
commutativity to constrain the class of processes that could possibly
be related to interfaces. Our proposal satisfies commutativity so they
can be viewed as a special case satisfying other relevant physical
constraints. Dub\'edat also came with a special solution of
commutativity. It corresponds to the case $m=0$ in our language. Then
the space of solutions has dimension $d_{n,0}=1$ and the corresponding
partition function is elementary:
\begin{equation}
\label{eq:dabdouwa}Z \propto \prod_{i <j} (x_j-x_i)^{2/\kappa}.
\end{equation}

\begin{figure}
\center{\includegraphics[width=0.5\textwidth]{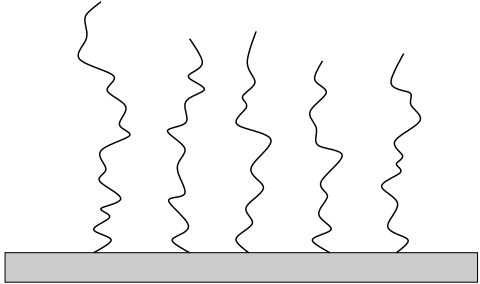}}
\caption{\emph{The factorisable $Z$ leads to a very simple geometry.
    This case has been suggested previously with a slightly different
    approach.}}
\label{fig: factorizable}
\end{figure}
A single arch topology is possible, all interfaces converge to
$\infty$, see figure \ref{fig: factorizable}.  Maybe this is a
good reason to call this case chordal $n$SLE.

\section{CFT background}

There was never any doubt that SLEs are related to conformal field
theories. The original approach
\cite{Bb:2002tf,Bb:2003vu,Bb:2003kd,Bb:2004ij} used the operator
formalism because if yields naturally martingale generating functions.
Here, we use the correlator approach for a change. We restrict the
presentation to a bare minimum, referring the newcomer to the many
articles, reviews and books on the subject (\cite{DMS:CFT,BPZ}). The
reader who knows too little or too much about CFT can profitably skip
this section.

Observables in CFT can be classified according to their behavior under
conformal maps. Local observables in quantum field theory are called
fields. For instance, in the Ising model, on an arbitrary (discrete)
domain, the average value of a product of spins on different (well
separated) sites can be considered. Taking the continuum limit at the
critical point, we expect that on arbitrary domains $D$ there is a
local observable, the spin. The product of two spins at nearest
neighbor points corresponds to the energy operator. In the continuum limit,
this will also lead to a local operator. In this limit, the lattice
spacing has disappeared and one can expect a definite (but
nontrivial) relationship between the energy operator and the product of two
spin fields close to each other. As on the lattice the product of two spins
at the same point is $1$, we can expect that the identity observable
also appears in such a product at short distances. Local fields come
in two types, bulk fields whose argument runs over $D$ and boundary
fields whose argument runs over $\partial D$. In this
paper, we shall not need bulk fields so we leave them aside.

The simplest conformal transformations in the upper-half plane are
real dilatations and boundary fields can be classified accordingly. It
is customary to write $\varphi_{\delta}(x)$ to indicate that in a real
dilatation by a factor $\lambda$ the field $\varphi_{\delta}(x)$ picks
a factor $\lambda^{\delta}$. By a locality argument, boundary fields
in a general domain $D$ (not invariant under dilatations) can still be
classified by the same quantum number. The number $\delta$ is called
the conformal weight of $\varphi_{\delta}$.

There are interesting
situations in which (due to degeneracies) the action of dilatations
cannot be diagonalized, leading to so called logarithmic CFT. While
this more general setting is likely to be relevant for several aspects
of SLE, we shall not need it in what follows. 

Under general conformal transformations, the simplest objects in CFT
are so called primary fields. Their behavior is dictated by the
simplest generalization of what happens under dilatations. Suppose
$\varphi_{\delta_1},\cdots \varphi_{\delta_n}$ are boundary primary
fields of weights $\delta_1,\cdots,\delta_n$. If $f$ is a conformal map
from domain $D$ to a domain $D'$, CFT postulates that
$$\bra 
\prod_{j=1}^n \varphi_{\delta_j}(x_j) \ket^D =  
 \bra
\prod_{j=1}^n \varphi_{\delta_j}(f(x_j)) \ket^{f(D)}
\prod_{j=1}^n |f'(x_j)|^{\delta_j}. 
$$
Symbolically, this can be written
$f:\varphi_{\delta}(x) \rightarrow
\varphi_{\delta}(f(x))|f'(x)|^{\delta}$.
It is interesting to make a comparison of these axioms with the
previous computations relating chordal SLE from $0$ to
$\infty$ to chordal SLE from $a$ to $b$ in several
normalizations. This also involved pure kinematics.

As usual in quantum field theory, to a symmetry corresponds an
observable implementing it. In CFT, this leads to the stress tensor
$T(z)$ whose conservation equation reduces to holomorphicity.  The
fact that conformal transformations are pure kinematics translates
into the fact that insertions of $T$ in known correlation functions
can be carried automatically, at least recursively. The behavior of
$T(z)$ under conformal transformations can be written as $f:T(z)
\rightarrow T(f(z))f'(z)^{2}+c/12Sf(z)$ where $Sf\equiv
(f''/f')'-1/2(f''/f')^2$ is the Schwarzian derivative and $c$ is a
conformal anomaly, a number which is the most important numerical
characteristic of a CFT. When $c=0$, $T$ is be a $(2,0)$ primary field
i.e. an holomorphic quadratic differential. When a (smooth) boundary
is present, the Schwarz reflection principle allows to extend $T$ by
holomorphicity.  Holomorphicity also implies that if $O$ is any local
(bulk or boundary) observable at point $z \in D$ and $v$ is vector
field meromorphic close to $z$, the contour integral $L_vO
\equiv\oint_z dw v(w)T(w)O$ along an infinitesimal contour around $z$
oriented counterclockwise is again a local field at $z$, corresponding
to the infinitesimal variation of $O$ under the map
$f(w)=w+\varepsilon v(w)$. It is customary to write $L_n$ for
$v(w)=w^{n+1}$. It is one of the postulates of CFT that all local
fields can be obtained as descendants of primaries, i.e. by applying
this construction recursively starting from primaries. The correlation
functions of descendant fields are obtained in a routine way from
correlations of the primaries. But descendant fields do not transform
homogeneously.

When $v$ is holomorphic at $x$, $L_vO$ is a familiar object. For
instance, if $\varphi_{\delta}$ is a primary boundary field, one
checks readily that $L_n\varphi_{\delta}=0$ for $n\geq 1$,
$L_0\varphi_{\delta}=\delta \varphi_{\delta}$ and
$L_{-1}\varphi_{\delta}=\Re e \; [\partial_x\varphi_{\delta}]$. The
other descendants are in general more involved, but by definition the
stress tensor $T=L_{-2}Id$ is the simplest descendant of the identity
$Id$. It does indeed not transform homogeneously.

A primary field and its descendants form what is called a conformal
family.  Not all linear combinations of primaries and descendants need
to be independent. The simplest example is the identity observable,
which is primary with weight $0$ and whose derivative along the boundary
vanishes identically\footnote{For other primary fields with the same
  weight if any, this does not have to be true.}. By contour
deformation, this leads to translation invariance of correlation
functions when $D$ has translation symmetry.

The next example in order of complexity is of utmost importance for
the rest of this paper. If $(2h+1)c=2h(8h-5)$, the field
$$-2(2h+1)L_{-2}\varphi_{h}+3L_{-1}^2\varphi_{h}$$
is again a primary, i.e. it transforms homogeneously under conformal
maps. In this case, consistent CFTs can be constructed for which it
vanishes identically. This puts further constraints on correlators. 

For example, when $D$ is the upper half plane, so that the Schwarz
principle extends $T$ to the full plane, the contour for $L_{-2}$ can be
deformed and shrunken at infinity.  Then, for an arbitrary boundary
primary correlator one has
\begin{eqnarray}
\label{eq:sing}
\left(\frac{3}{2(2h+1)}\partial_x^2 +\sum_{\alpha=1}^{l}\left[
  \frac{1}{y_{\alpha}-x}\partial_{y_{\alpha}}-
  \frac{\delta_{\alpha}}{(y_{\alpha}-x)^2} \right]\right) & & \nonumber \\
& & \hspace{-3cm}  \bra \varphi_{\delta}(\infty) \prod_{\alpha=1}^l
\varphi_{\delta_{\alpha}}(y_{\alpha}) \varphi_{h}(x)
\ket=0. \end{eqnarray}  
It is customary to call this type of equation a null-vector equation.

Note that the primary field of weight $\delta$ sitting at $\infty$ has
led to no contribution in this differential equation. Working the
other way round, this equation valid for an arbitrary number of
boundary primary fields with arbitrary weights characterizes the field
$\varphi_{h}$ and the relation between $h$ and the central charge $c$.

The case of three points correlators is instructive.  Global conformal
invariance implies that
$$\bra \varphi_{\delta}(y)\varphi_{\delta'}(y') \varphi_{h}(x) \ket
\propto |y-y'|^{h-\delta-\delta'}
|x-y|^{\delta'-h-\delta}|y'-x|^{\delta-\delta'-h}.$$
The
proportionality constant might depend on the cyclic ordering of the
three points. But if the differential equation for $\varphi_{h}$ is
used, a further constraint appears. The three point function can be
non vanishing only if
$$
3(\delta-\delta')^2-(2h+1)(\delta+\delta')= h(h-1).$$
This
computation has a dual interpretation : consider a correlation
function with any number of fields, among them a $\varphi_{\delta}(y)$
and a $\varphi_{h}(x)$. If $x$ and $y$ come very close to each other
they can be replaced by an expansion in terms of local fields. This is
called fusion. Several conformal families can appear in such an
expansion, but within a conformal family, the most singular
contribution is always from a primary. This argument applies even if
$c$ and $h$ are arbitrary.  But suppose they are related as above and
the differential equation eq.(\ref{eq:sing}) is valid. This equation
is singular at $x=y$ and at leading order the dominant balance leads
to an equation where the other points are spectators. One finds that
the only conformal families that can appear are the ones whose
conformal weight $\delta'$ satisfies the fusion rule.

This is enough CFT background for the rest of this paper. We are now
in position to give the heuristic argument that leads to our main claims.

\section{Martingales from statistical mechanics}
\label{sec:StatMech}

The purpose of this section is to emphasize the intimate connection
between the basic rules of statistical mechanics and martingales. The
connection is somehow tautological, because statistical mechanics
works with partition functions, i.e. unnormalized probability
distributions, all the time. In the discrete setting, this makes
conditional expectations a totally transparent operation that one
performs without thinking and even without giving it a name. But the
following argument is, despite its simplicity and its abstract
nonsense flavor, the crucial observation that allows us to relate
interfaces in conformally invariant statistical mechanics to SLEs.

\subsection{Tautological martingales}

Consider a model of statistical mechanics with a finite but
arbitrarily large set of possible states $S$. Usually one starts with
models defined on finite grid domains so $\# S < \infty$ is natural.
To each state $s \in S$ we associate a Boltzmann
weight\footnote{Usually the Boltzmann weight is related to the energy
  $H(s)$ of the state $s$ through $w(s) = \exp ( - \beta H(s))$, where
  $\beta$ is the inverse temperature (a Lagrangian multiplier related
  to temperature, anyway).}  $w(s)$. The partition function is $Z =
\sum_{s \in S} w(s)$ so that it normalizes the Boltzmann weights to
probabilities, $\prob \{ s \} = \frac{w(s)}{Z}$.  Since $S$ is finite,
we can use the power set $\sP (S) = \{ U : U \subset S \}$ as a sigma
algebra.  The expected value of a random variable $\Oper : S
\rightarrow \bC$ is denoted by $\expect [\Oper] = \bra \Oper \ket =
\frac{1}{Z} \sum_{s \in S} \Oper(s) w(s)$.

Note that if $(S_\alpha)_{\alpha \in I}$ is a collection of disjoint
subsets of $S$ such that $\cup_{\alpha \in I} S_\alpha = S$, then the
collection of all unions $\sF = \{ \cup_{\alpha \in I'} S_\alpha : I'
\subset I \}$ is a sigma algebra on $S$. Conversely, since $S$ is
finite, any sigma algebra $\sF$ on $S$ is of this type.

Consider a filtration, that is an increasing family $(\sF_t)_{t \geq
  0}$ of sigma algebras $\{ \emptyset , S \} \subset \sF_s \subset
\sF_t \subset \sP(S)$ for all $0 \leq s < t$.  Denote the
corresponding collections of disjoint sets by
$(S^{(t)}_{\alpha})_{\alpha \in I_t}$ and define the partial partition
function $Z^{(t)}_{\alpha}\equiv \sum_{s \in S^{(t)}_\alpha} w(s)$.
The conditional expectation values
\begin{eqnarray*}
 \bra \Oper \ket_t & \equiv  & \expect [ \Oper | \sF_t]=
    \sum_{\alpha \in I_t} \frac{\sum_{s \in S^{(t)}_\alpha} 
    \Oper(s) w(s)}
    {\sum_{s \in S^{(t)}_\alpha} w(s)} \; \unit_{S^{(t)}_\alpha} \\
& = & \sum_{\alpha \in I_t} \Big( \frac{1}{Z^{(t)}_{\alpha}}
    \sum_{s \in S^{(t)}_\alpha} \Oper(s) w(s) \Big) \;
    \unit_{S^{(t)}_\alpha} 
\end{eqnarray*}
are martingales by definition: for $s < t$ we have
\begin{eqnarray*}
\expect \big[ \; \expect [ \Oper | \sF_t] \; \big| \sF_s \big]
    = \expect [ \Oper | \sF_s]
\end{eqnarray*}
Notice that the probability of the event $S^{(t)}_\alpha$ is
conveniently $\prob [S^{(t)}_{\alpha}] = Z^{(t)}_\alpha / Z$.

\begin{figure}
\center{\includegraphics[width=0.6\textwidth]{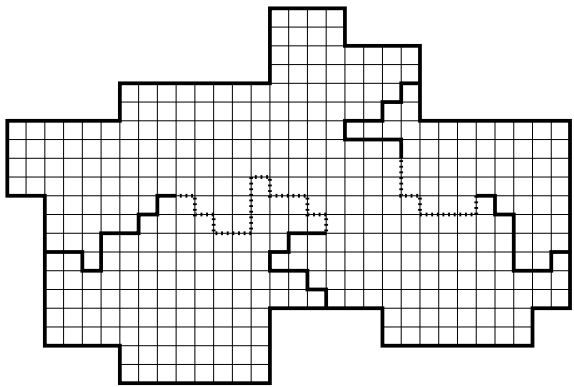}}
\caption{\emph{A discrete statistical mechanics model with
portions of interfaces specified.}}
\label{fig:statmech}
\end{figure}

Suppose that the model is defined in a domain $D \subset \bC$ and that
there are interfaces in the model. Parametrize portions of these
interfaces touching the boundary by an arbitrary ``time'' parameter
$t$ in such a way that $n$ paths $\gamma^{(i)}_t$, $i=1,\cdots,n$
(which are pieces of the random interfaces) emerge from the boundary
at $t=0$ and are disjoint at least when $t$ is small enough, see
figure \ref{fig:statmech}. To avoid
confusion we write the time parameter $t$ now as a subscript and
continue to indicate the dependence of $s \in S$ by parenthesis, so $t
\mapsto \gamma^{(i)}_t (s)$ is a parametrization of the $i^{\textrm{th}}$ piece
of interface if the system is at state $s$.  Then we can consider the
natural filtration of the interface by taking $\sF_t = \sigma (
\gamma^{(i)}_{t'} : 0 \leq t' \leq t, \; i=1,\cdots,n )$ to be the
sigma algebra generated by the random variables $\gamma^{(i)}_{t'}$ up
to time $t$.

The boundary conditions of the model are often
such that conditioning on the 
$\gamma^{(i)}_{[0,t]}$, is the same as considering the model in a
smaller domain (a part of the interface removed) but with
same type of boundary conditions. Of course the position at which
the new interface should start is where the original interface
would have continued, that is the $\gamma^{(i)}_t$'s. Let $D_t$ be the
domain $D$ with the $\gamma^{(i)}_{]0,t]}$ removed. 

The starting point of the next section is the input of conformal
invariance in this setup.

\subsection{Simplifying tautological martingales}

We start from the situation and notations at the end of the previous
section.
If in addition we are considering a model at
its critical point, then the continuum limit may be described
by a conformal field theory. At least for a wide class of natural
observables $\Oper$, the expectation values become CFT correlation
functions in the domain $D$ of the model
\begin{eqnarray*}
\bra \Oper \ket = \frac{\sum_{s \in S} \Oper(s) w(s)}{Z}
\longrightarrow \frac{\bra \Oper \ket^{\textrm{CFT, b.c.}}_{D}}
{\bra \unit \ket^{\textrm{CFT, b.c.}}_{D}}
\end{eqnarray*}
We need to write the correlation function of identity (proportional to
$Z$) in the denominator because the boundary conditions (b.c.)  of the
model may already have led to insertions of boundary changing
operators that we have not mentioned explicitly. 

The closed martingales become
\begin{eqnarray*}
\bra \Oper \ket_t = \sum_{\alpha \in I_t} \frac{1}{Z^{(t)}_\alpha}
    \sum_{s \in S^{(t)}_\alpha} \Oper(s) w(s) \; \unit_{S^{(t)}_\alpha}
\longrightarrow
\frac{\bra \Oper \ket^{\textrm{CFT, b.c.}}_{D_t}}
    {\bra \unit \ket^{\textrm{CFT, b.c.}}_{D_t}}
\end{eqnarray*}
where in the continuum limit $D_t$ might be $D$ with hulls (and not
only traces) removed.

For certain (but not all) observables, $\bra \Oper \ket$ is computing
a probability, which in a conformal field theory ought to be
conformally invariant.  But $\bra \Oper \ket$ is written as a quotient,
and this means that the numerator and denominator should transform
homogeneously (and with the same factor) under conformal
transformations. In particular, the denominator should transform
homogeneously. This means that $\bra \unit \ket^{\textrm{CFT,
    b.c.}}_{D}$ -- which depends on the position of the boundary
condition changes -- behaves like a product of boundary primary fields.
Then, by locality, for any $\Oper$, the transformation of the
numerator under conformal maps will split in two pieces: one
containing the transformations of $\Oper$ and the other one
canceling with the factor in the denominator.  So we infer the
existence in the CFT of a primary boundary field, denoted by $\psi (x)$
in what follows, which implements boundary condition changes at which
interfaces anchor.  Hence we may write 
$$\bra \unit \ket^{\textrm{CFT, b.c.}}_{D} = \bra \psi (X^{(1)})
\cdots \psi (X^{(n)})\ket^{\textrm{CFT}}_{D}$$
and
$$\bra \Oper \ket^{\textrm{CFT, b.c.}}_{D} = \bra \Oper \psi (X^{(1)})
\cdots \psi (X^{(n)})\ket^{\textrm{CFT}}_{D}.$$

As will become clear later, there might also be one further boundary operator
anchoring several interfaces. We do not mention it explicitly here
because it will sit at a point which will not be affected
by the conformal transformations that we use. 

Write the transformation of the observable $\Oper$ as $f:\Oper
\rightarrow \; ^f \!\Oper$ under a conformal map. Denote by $f_t$ a
conformal representation $f_t:D_t\rightarrow D$ and write
$f(\gamma^{(i)}_{t}) \equiv X^{(i)}_{t}$. The expression for the closed
martingale $\bra \Oper \ket_t$ can now be simplified further
\begin{equation}
\label{eq:martCFT}
\bra \Oper \ket_t \longrightarrow \frac{\bra \, ^{f_t}\Oper\, \psi
  (X^{(1)}_t) \cdots \psi (X^{(1)}_t)\ket^{\textrm{CFT}}_{D}} {\bra
  \psi (X^{(1)}_t) \cdots \psi (X^{(1)}_t)
  \ket^{\textrm{CFT}}_{D}}.
\end{equation}
The Jacobians coming from the transformations of the boundary changing
primary field $\psi$ have canceled in the numerator and denominator.
The explicit value of the conformal weight of $\psi$ does not appear
in this formula. 

Of course, we have cheated. For the actual map $f_t$ which is
singular at the $\gamma^{(i)}_{t}$'s these Jacobians are infinite. A
more proper ``derivation'' would go through a regularization but
locality should ensure that the naive formula remains valid when the
regularization is removed. Eq.(\ref{eq:martCFT}) is the starting
point of our analysis.

\section{Derivation of the proposal}

The heuristics we follow is to describe a growth process of interfaces
by a Loewner chain $f_t$ compatible with conformal invariance in that
the right hand side of eq.(\ref{eq:martCFT}) is a martingale.

\subsection{The three ingredients}

\noindent \textbf{Loewner chain}: If we use the upper half plane as a
domain, $D={\mathbb H}$,  and 
impose the hydrodynamic normalization, the equation for $f_t$ has to
be of the form
$$\ud f_t(z)=\sum_i \frac{2\ud q^{(i)}_t}{f_t(z)-X^{(i)}_t}$$
for some processes $X^{(i)}_t$, $i=1,\ldots ,n$. The initial condition
is $f_0(z)=z$.

\vspace{.3cm}

\noindent \textbf{Interfaces grow independently of each other on very
  short time scales}: Schramm's argument deals with the case of a
single point. We expect that on very short time scales the growth
processes do not feel each other and Schramm's argument is still
valid, so that $\ud X^{(i)}_t=\ud M^{(i)}_t + F^{(i)}_t $ where the
$M^{(i)}_t$'s are $n$ (continuous,local) martingales with quadratic variation
$\kappa q^{(i)}_t$ and vanishing cross variation and $F^{(i)}_t$ is a
drift term.

\vspace{.3cm}

\noindent \textbf{The martingale property fixes the drift term}: The
drift term will be computed by imposing the martingale condition on
$\bra \Oper \ket_t$ when $\Oper$ is a product of an arbitrary number
$l$ of boundary primary fields $\Oper =\prod_{\alpha=1}^{l}
\varphi_{\delta_\alpha}(Y^{(\alpha)})$. The insertion points are away
from the boundary changing operators and $f_t$ is regular with
positive derivative there.  Substitution of $^{f_t}\Oper$ in formula
(\ref{eq:martCFT}) yields
\begin{equation}
\label{eq:marto}
 \bra \prod_{\alpha=1}^{l}
  \varphi_{\delta_\alpha}(Y^{(\alpha)})\ket_t = \frac{\bra
    \prod_{\alpha=1}^{l} 
  \varphi_{\delta_\alpha}(f_t(Y^{(\alpha)})) 
  \prod_{i=1}^{n}\psi (X^{(i)}_t) \ket^{\textrm{CFT}}_{D}} 
{\bra \prod_{i=1}^{n}\psi (X^{(i)}_t) \ket^{\textrm{CFT}}_{D}}
\prod_{\alpha=1}^{l} f'_t(Y^{(\alpha)})^{\delta_\alpha}. 
\end{equation}

\subsection{Computation of the Ito derivative of $\bra
  \prod_{\alpha=1}^{l} \varphi_{\delta_\alpha}(Y^{(\alpha)})\ket_t$}

In formula (\ref{eq:marto}), denote respectively by $Z_t^{\varphi}$,
$Z_t$ and $J^{\varphi}_t$ the numerator, denominator and Jacobian
factor on the right hand side.

It is useful to break the computation of $\ud \bra
\prod_{\alpha=1}^{l} \varphi_{\delta_\alpha}(Y^{(\alpha)})\ket_t$ in
several steps.

\vspace{.3cm}

-- \textit{Preliminaries}.  \\
Ito's formula for the $\psi$'s gives
$$\ud \psi(X^{(i)}_t)=\psi'(X^{(i)}_t)(\ud M^{(i)}_t +
F^{(i)}_t)+\frac{\kappa}{2} \psi''(X^{(i)}_t)\ud q^{(i)}_t. $$
\noindent Using
the Loewner chain for $f_t(z)$ and its derivative with respect to $z$,
one checks that
$$\ud
\left(\varphi_{\delta}(f_t(Y))f'_t(Y)^\delta\right)=f'_t(Y)^\delta
\sum_i 2\ud
q^{(i)}_t\left(\frac{\varphi'_{\delta}(f_t(Y))}{f_t(Y)-X^{(i)}_t}-
  \frac{\delta\varphi_{\delta}(f_t(Y))}{(f_t(Y)-X^{(i)}_t)^2}\right).$$

\vspace{.2cm}

-- \textit{The Ito derivative of $Z_t^{\varphi}J^{\varphi}_t$}.\\
The time $t$ being given, we can simplify the notation. Set $x_i \equiv
X^{(i)}_t$ and $y_{\alpha} \equiv f_t(Y^{(\alpha)})$ and apply the
chain rule to get 
\begin{eqnarray*}
\frac{\ud (Z_t^{\varphi}J^{\varphi}_t)}{J^{\varphi}_t} & = &
\left[\sum_i \left(\ud M^{(i)}_t + 
F^{(i)}_t\right)\partial_{x_i}\right. \\
& &  \hspace{-2cm} + \left. \sum_i \ud q^{(i)}_t\left(\frac{\kappa}{2} 
\partial_{x_i}^2+2\sum_{\alpha}\left[
\frac{1}{y_{\alpha}-x_i}
\partial_{y_{\alpha}}-\frac{\delta_{\alpha}}
{(y_{\alpha}-x_i)^2}\right]\right)\right]Z_t^{\varphi}
\end{eqnarray*}

\vspace{.2cm}

-- \textit{First use of the null-vector equation : identification
of $\psi$}. \\
Let us concentrate for a moment on the familiar chordal SLE case, for
which $n=1$. The drift term $F_t^{(1)}$ is known to be zero. The
boundary conditions also change at $\infty$ (the endpoint of the
interface) and there is an operator there, that we have not written
explicitly because the notation is heavy enough. Anyway, $Z_t$ is a
two-point function with one of the fields at infinity, so it is a
constant. For chordal SLE, the drift term in the Ito derivative of the
putative martingale vanishes if and only if
$$\left(\frac{\kappa}{2} 
\partial_{x}^2+2\sum_{\alpha}\left[
\frac{1}{y_\alpha-x}
\partial_{y_\alpha}-\frac{\delta_{\alpha}}
{(y_\alpha-x)^2}\right]\right)Z_t^{\varphi}=0,$$
where for simplicity we have written $x \equiv x_1$.\\
Comparison with eq.(\ref{eq:sing}) implies that $\psi$ has a vanishing
descendant at level two and has conformal weight
$h=\frac{6-\kappa}{2\kappa}\equiv h_1(\kappa)$ : 
$$\psi (x)\equiv \varphi_{h_1(\kappa)}(x).$$ 
The central charge is
$c=\frac{(6-\kappa)(3\kappa-8)}{16\kappa}$. \\
This is of course nothing but the correlation function formalism
version of the original argument relating SLE to CFT, which was given
in the operator formalism, see \cite{Bb:2002tf}.

\vspace{.2cm}

-- \textit{Second use of the null-vector equation}.\\
Now that $\psi$ has been identified, we can return to the general
case, with an arbitrary number $n$ of growing curves. Each growing
curve has its own field $\psi$ and each field $\psi$ comes with its
differential equation, which is eq.(\ref{eq:sing}) but for $l+n-1$
spectator fields, the $l$ fields $\varphi$ and the $n-1$ other
$\psi$'s. So $Z_t^{\varphi}$ is annihilated by the $n$ differential
operators
$$
\frac{\kappa}{2} 
\partial_{x_i}^2+2\sum_{\alpha}
\left[ \frac{1}{y_\alpha-x_i}
\partial_{y_\alpha}-\frac{\delta_{\alpha}}
{(y_\alpha-x_i)^2}\right] +2\sum_{j \neq i}\left[
\frac{1}{x_j-x_i}
\partial_{x_j}-\frac{h_1(\kappa)}
{(x_j-x_i)^2}\right].
$$
We can use this to get a simplified formula
$$
\ud (Z_t^{\varphi}J^{\varphi}_t)= J^{\varphi}_t \, {\mathcal
  P}Z_t^{\varphi}\; , \qquad \ud Z_t={\mathcal P} Z_t$$
where ${\mathcal P}$ is the first order differential operator
$$
\sum_i \left[
\left(\ud M^{(i)}_t + F^{(i)}_t\right)\partial_{x_i}- 2\ud q^{(i)}_t
\left(\sum_{j \neq i} \left[\frac{1}{x_j-x_i}\partial_{x_j}
-\frac{h_1(\kappa)}{(x_j-x_i)^2}\right] \right)\right].
$$
The formula for $Z_t$ is just the special case $l=0$. 

\vspace{.2cm}

-- \textit{Final application of Ito's formula}.\\
$$
\ud \left(\frac{Z_t^{\varphi}}{Z_t}J^{\varphi}_t
\right)=J^{\varphi}_t {\mathcal
  Q}\left(\frac{Z_t^{\varphi}}{Z_t}\right)
$$
where ${\mathcal Q}$ is the first order differential operator
$$\sum_i \left[\ud M^{(i)}_t + F^{(i)}_t- \kappa \ud
  q^{(i)}_t(\partial_{x_i}\log Z_t) - 2 \sum_{j \neq i} \frac{\ud
    q^{(j)}_t}{x_i-x_j}\right]\partial_{x_i}$$
The martingale property is satisfied if and only if the drift terms vanish.

\subsection{Main claim}

To summarize, we have shown that the system 
$$\ud f_t(z)=\sum_i \frac{2\ud q^{(i)}_t}{f_t(z)-X^{(i)}_t} \quad ,
\quad \ud X^{(i)}_t=\ud M^{(i)}_t + F^{(i)}_t$$
admits conditioned
correlation functions from CFT as martingales if and only if
$$F^{(i)}_t=\kappa \ud
  q^{(i)}_t(\partial_{x_i}\log Z_t) + 2 \sum_{j \neq i} \frac{\ud
    q^{(j)}_t}{x_i-x_j}.$$
where $Z_t$ is a partition function.
It is under this condition that it describes the growth of $n$
interfaces in a way compatible with statistical mechanics and
conformal field theory. 

\vspace{.3cm}

In fact, we have used a special family of correlators. But the same
argument applies to all operators (hence the ``if'' part). Of special
interest in the sequel will be the case when $\Oper$ is a topological
observable, for instance taking value $1$ if the interface forms a
given arch system and $0$ otherwise. No Jacobian is involved for such
observables and the numerator looks again like a partition function. 

\subsection{The moduli space} 

From the definition of $Z_t$ as a correlation of primary fields with
null descendants at level $2$, it is clear that properties
$i),\;ii),\;iii)$ are satisfied, except maybe for the quantization of
the possible scaling dimensions of $Z_t$, to which we turn now.

This is standard material from CFT and we include it here for
completeness.

The correlator $\bra \varphi_{h_{\infty}} (\infty)\psi(x_1)
\cdots \psi(x_n)\ket$ on the real line satisfies $n$
differential equations. We shall recall why the space of simultaneous
solutions which have global conformal invariance has dimension
$${n \choose m}-{n \choose m-1}=(n+1-2m)\frac{n!}{m!(n-m+1)!}$$
if $h_{\infty}= h_{n-2m}(\kappa)$ for some nonnegative integer
$m \leq n/2$ and has dimension $0$ otherwise. This will end
the derivation of our proposal and match the counting of arches.

At the end of the background on conformal invariance, we mentioned
fusion rules: when $\varphi_{h_{1}(\kappa)}(x)$ and a
$\varphi_{h_{j}(\kappa)}(y)$ are brought close together, they can be
expanded in a basis of local operators that can be grouped in
conformal families. We also recalled why the weight $h'$ of the
primaries in each conformal family had to satisfy
$3(h_{j}(\kappa)-h')^2-(2h_{1}(\kappa)+1)(h_{j}(\kappa)+h')=
h_{1}(\kappa)(h_{1}(\kappa)-1),$ so that only two conformal families
can appear in a fusion with $\varphi_{h_{1}(\kappa)}$. The two
conformal weights  are easily found to
be $h'=h_{j\pm 1}(\kappa)$. Furthermore, $h_0(\kappa)=0$ and one can
show that the corresponding field has to be the boundary identity
operator. By global conformal invariance, the only local operator with
a non vanishing one point correlator is the identity and boundary two
point functions vanish unless the two local fields have the same
conformal weight. This takes care of the counting and selection rules
for the $n=0,1$ cases.

One proceeds by recursion.  The points are ordered $x_1< x_2 \cdots <
x_n$. If $n\geq 2$ then move $x_2$ close the $x_1$ (for instance by a
global conformal transformation) and fuse to get an expansion for
local fields at $x_1$ say. Only the conformal families of
$\varphi_{h_{1\pm 1}(\kappa)}$ appear. If $n=2$ this fixes the weight
of the field at $\infty$. If $n \geq 3$, iterate. This leads
immediately to the selection rules mentioned above : the field at
infinity has to be a $\varphi_{h_{n-2m}(\kappa)}$. The dimension is
nothing but the number of path of $n$ steps $\pm 1$ from $0$ to $n-2m$
on the nonnegative integers, a standard combinatorial problem whose
answer is ${n \choose m}-{n \choose m-1}$. The efficient way to do the
counting is by the reflection principle. The possible outcomes of each
fusion can be encoded in a so-called Bratelli diagram:
\begin{equation}
\label{eq: fusion diagram}
\left. \begin{array}{ccccccccccc}
       &          &                &          &                 &          &                 &          &                 &          & \cdots\\
      &          &                &          &                 &          &                 &          &                 & \nearrow & \\
      &          &                &          &                 &          &                 &          & h_{4}(\kappa)  &          & \\
      &          &                &          &                 &          &                 & \nearrow &                 & \searrow & \\
      &          &                &          &                 &          & h_{3}(\kappa)   &          &                 &          & \cdots \\
      &          &                &          &                 & \nearrow &                 & \searrow &                 & \nearrow & \\
      &          &                &          & h_{2}(\kappa)   &          &                 &          & h_{2}(\kappa)      &          & \\
      &          &                & \nearrow &                 & \searrow &                 & \nearrow &                 & \searrow & \\
      &          & h_{1}(\kappa)  &          &                 &          & h_{1}(\kappa)   &          &                 &          & \cdots \\
      & \nearrow &                & \searrow &                 & \nearrow &                 & \searrow &                 & \nearrow & \\
h_{0}(\kappa) &          &                &          & h_{0}(\kappa)   &          &                 &          & h_{0}(\kappa)      &          & \\
      &          & \quad          &          & \quad           &          & \quad           &          & \quad           &          & \\
      &          & \textrm{$1$SLE}&          & \textrm{$2$SLE} &          & \textrm{$3$SLE} &          & \textrm{$4$SLE} &          & \cdots \\
\end{array} \right.
\end{equation}

This is totally parallel to the discussion of composition of $n$ spins
$1/2$ for the representation theory of the Lie algebra of rotations.
The multiplicity is exactly one when $m=0$ which corresponds to the
partition function (\ref{eq:dabdouwa}) and to the insertion of the
operator $\varphi_{h_{n}(\kappa)}$ at infinity, toward which the $n$
interfaces run.

What is not proved here is
that the different paths lead to a basis of solutions of the $n$
partial differential equations, but it is true. Each path corresponds
to a succession of choices of a single conformal family, one at each
fusion step. Let us mention in advance that multi SLE processes, i.e.
the consideration of multiple interfaces, will lead to the definition
of another basis with a topological interpretation.

\section{Multiple SLEs describing several interfaces}
\label{sec:sevinterf}

\subsection{Double SLEs}
\label{sec: double SLE}
The case of double SLEs is instructive and simple to analyze.
Although double SLEs is sometimes interesting for its own sake,
the purpose of this section is to give easy examples to guide
the study of the general case.

\subsubsection{2SLEs and  Bessel processes}
To specify the process we have to specify the partition function $Z$.
There are only two possible choices corresponding to two different
type of boundary conditions, or alternatively to two different fields
inserted at infinity: 
\begin{eqnarray*}
\bra h_\infty|\psi(X_1)\psi(X_2) | 0 \ket
& = & \const \times (X_1 - X_2)^{\Delta}
\end{eqnarray*}
where the exponent is $\Delta = h_\infty - 2 h_{1}(\kappa)$ and the
constant will be fixed to $1$ from now on.  According to CFT fusion
rules, $h_\infty$ can only be either $h_{2}(\kappa) =
\frac{8-\kappa}{\kappa}$ or $h_{0}(\kappa)=0$. The exponent becomes
$\Delta = 2/\kappa$ or $\Delta = \frac{\kappa-6}{\kappa}$
respectively, so that we have two basic choices for $Z$:
$$
Z_0\equiv (X_1-X_2)^{(\kappa-6)/\kappa}\quad {\rm or}\quad
Z_2 \equiv (X_1-X_2)^{2/\kappa}
$$
As we shall see, choosing $Z_0$ selects configurations with no
curve ending at infinity -- so that we are actually describing
standard chordal SLE joining to the two initial positions of $X_1$ and
$X_2$ -- while choosing $Z_2$ selects configurations with two curves
emerging from the initial positions of $X_1$ and $X_2$ and ending both
at infinity.

Up to normalizing the quadratic variation by $dq^{(i)}_t= a_i dt$ so
that the martingales $M^{(i)}$ are simply 
$dM^{(i)}_t=\sqrt{\kappa a_i}dB_t^{(i)}$ 
with $dB_t^{(i)}$ two independent normalized Brownian motions,
our double SLE equations become~:
\begin{eqnarray*}
df_t(z) & = & \frac{2 a_1\;\ud t}{f_t(z) - X^{(1)}_t}
    + \frac{2 a_2\;\ud t}{f_t(z) - X^{(2)}_t} \\
d X^{(1)}_t & = & \sqrt{a_1 \kappa} \; d B^{(1)}_t +
    \frac{2 a_2 + \kappa \Delta a_1}{X^{(1)}_t - X^{(2)}_t} \; \ud t \\
d X^{(2)}_t & = & \sqrt{a_2 \kappa} \; d B^{(2)}_t +
    \frac{2 a_1 + \kappa \Delta a_2}{X^{(2)}_t - X^{(1)}_t} \; \ud t
\end{eqnarray*}
It describes two curves emerging from points $X_1= X^{(1)}_0$ and
$X_2=X^{(2)}_0$ at speeds parametrized by $a_1$ and $a_2$.

Up to an irrelevant translation, the process is actually driven by the
difference $Y_t=X^{(1)}_t-X^{(2)}_t$.  Up to a time change, $ds =
\kappa (a_1 + a_2)dt$, this is a Bessel process,
\begin{eqnarray*}
dY_s = \ud \tilde{B}_s +
    \frac{\Delta + 2/\kappa}{Y_s} \; \ud s,
\end{eqnarray*}
of effective dimension $d_{\rm eff}=1 + 2\Delta + 4/\kappa$. For
$h_\infty = h_{2}(\kappa)$ (i.e. $\Delta=2/\kappa$) the dimension is 
$d_{\rm eff}=1+ 8/\kappa$ and for $h_\infty = 0$
(i.e. $\Delta=(\kappa-6)/\kappa$) it is $d_{\rm eff}=3 - 8/\kappa$.
In the physically interesting parameter range $\kappa < 8$,
the former is $> 2$ and the latter is $< 2$. 
Recall now that a Bessel process is recurrent (not recurrent) if its
effective dimension is less (greater) than $2$. 
Thus, the driving processes $X^{(i)}_t$ hit each other almost
surely in the case $h_{\infty} = 0$ and they don't hit (a.s.)
in the case $h_\infty =h_{2}(\kappa)$. Since the hitting of driving
processes means hitting of the SLE traces, this teaches us that case
$h_\infty=0$ describes a single curve joining $X_1$ and $X_2$ while case
$h_\infty=h_{2}(\kappa)$ describes two curves converging toward infinity.

Notice that previous results are independent of $a_1$ and $a_2$,
provided their sum does not vanishes.
We also observe that setting $a_1 = 1$ and $a_2 = 0$ (or vice versa)
one recovers an SLE$(\kappa; \kappa \Delta)$. Recall that if
$h_\infty = 0$ then $\rho = \kappa \Delta = \kappa - 6$ corresponds
to an ordinary chordal SLE from $X_1$ to $X_2$. Our
double SLEs with $h_\infty = 0$ corresponds to one chordal SLE seen
from both ends and the fact that the tips of the traces hit is
natural. The other case, $h_\infty = h_{2}(\kappa)$ corresponds to
$\rho = \kappa \Delta = 2$ and since the driving processes can not
hit, the process can be defined for all $t \geq 0$.  Assuming
that $\int_0^\infty ( a_1 + a_2) \ud t = \infty$, the
capacity of the hulls grow indefinitely and (at least one of) the
SLE traces go to infinity.

\begin{figure}
\includegraphics[width=1.0\textwidth]{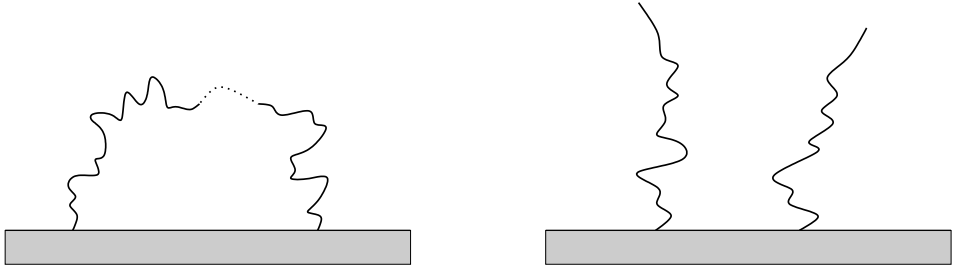}
\caption{\emph{The two geometries for 2SLE: on the left is the case
$h_\infty = 0$ and on the right $h_\infty = h_2 (\kappa)$.}}
\label{fig: 2SLEs}
\end{figure}

The two possible geometries are illustrated in figure \ref{fig: 2SLEs}.

\subsubsection{A mixed case for 2SLE}
\label{sec: mixed 2SLE}
Because of its simplicity, we use double SLE as a testing ground for
mixed correlation functions. So we consider the sum
$$ Z = \lambda Z_0 + \mu Z_2$$ 
with both $\lambda$ and $\mu$ positive. As already mentioned, the
interpretation of $Z$ as the continuum limit of partition functions of
lattice models is unclear since $Z_0$ and $Z_2$ do not scale the same way.
We nevertheless study it to illustrate ways of
computing (arch or geometry) probabilities.
As one may expect, we no longer have an almost sure global geometry
but rather nontrivial probabilities for the two geometries: either no
curve at infinity or two curves converging there.

Let $\tau = \inf \{ t \geq 0 : X_t^{(1)} = X_t^{(2)} \}$ be the
stopping time which indicates the hitting of the driving processes --
and thus of the two curves.  We can define the driving processes as
solutions of the 2SLE system on the (random) time interval $t \in [0,
\tau)$.  At the stopping time we define $f_\tau(z) = \lim_{s \uparrow
  \tau} f_s(z)$ for such $z \in \bH$ that the limit exists and stays
in the half plane $\bH$. The hull $K_\tau$ is defined as the set where
the limit doesn't exist or hits $\bdry \bH$.

The question of geometry is answered by the knowledge of whether
the two traces hit, that is whether $\tau < \infty$ or not.
Thus we again consider the difference
$Y_t = X^{(1)}_t - X^{(2)}_t$, whose Ito derivative is now~:
\begin{eqnarray*}
\ud Y_t & = &  \sqrt{\kappa} \ud \tilde B_t + \frac{2}{Y_t}(a_1+a_2)\ud t
+ \frac{(\kappa - 6) \lambda Y_t^{\frac{\kappa-6}{\kappa}} + 
2 \mu Y_t^{\frac{2}{\kappa}}}{Y_t(\lambda Y_t^{\frac{\kappa-6}{\kappa}} + 
\mu Y_t^{\frac{2}{\kappa}} )} \; (a_1+a_2)\ud t
\end{eqnarray*}
with $\tilde B_t=\sqrt{a_1}B^{(1)}_t-\sqrt{a_2}B^{(2)}_t$ is a
Brownian motion, so that after a time change, $\ud s= (a_1+a_2)\ud t$,
the result doesn't depend on $a_1$ or $a_2$.  The last drift term
comes from the derivative of $\log Z$.
 
One might for example try to find the distribution of $\tau$ by its
Laplace transform $\expect_{Y_0 = y} [ e^{-\beta \tau} ] =
f_\beta(y)$. By Markov property,
\begin{eqnarray*}
\expect [ e^{-\beta \tau} | \sF_t] = e^{-\beta t} f_\beta (Y_t)
\end{eqnarray*}
is a closed martingale on $t \in [0, \tau)$ so requiring its Ito drift
to vanish leads to the differential equation
\begin{eqnarray*}
\Big( -\frac{\beta}{a_1 + a_2} + \big( \frac{2}{y} + \frac{(\kappa-6)
    \lambda + 2 \mu y^{(8-\kappa)/\kappa}}{\lambda 
    + \mu y^{(8-\kappa)/\kappa}} \big) \partial_y
    + \frac{\kappa}{2} \partial_y^2 \Big) f_\beta (y) = 0
\end{eqnarray*}
The result depends only on $\beta / (a_1 + a_2)$. We conclude that
the distribution of $(a_1 + a_2) \tau$,
the capacity of the final hull $K_\tau$,
is independent of the speeds of
growth $a_1$ and $a_2$. Also the result depends on $\lambda$ and
$\mu$ only through $\mu / \lambda$.

In particular we want to take $\beta \downarrow 0$ to compute
the probability that the traces hit. Constant functions solve the
differential equation but another linearly independent solution
has the correct boundary values $f_0 (0)=1$ and $f_0 (\infty)=0$,
namely
\begin{eqnarray*}
\prob_{Y_0 = y} [\tau < \infty] = \lim_{\beta \downarrow 0}
    \expect_{Y_0 = y} [e^{-\beta \tau}]
= \frac{\lambda}{\lambda+{\mu} y^{(8-\kappa)/\kappa}}
\end{eqnarray*}
As expected on general ground, this is the fraction of the two
partition functions $\lambda Z_0$ and $Z = \lambda Z_0 + \mu Z_2$.

\subsection{Triple and/or quadruple SLEs}
We will give a few more of examples of multiple SLEs.
Certain triple and quadruple SLEs are the scaling limits of
interfaces in percolation and Ising model with rather natural
boundary conditions. These models will be considered in
section \ref{sec: Ising}. Here we study triple and quadruple
SLEs for their own sake. We restrict ourselves to $\kappa<8$.

\subsubsection{3SLE (pure) configurations}
Partition functions with $n=3$ have only two possible scaling
behaviors depending whether the weight $h_\infty$ of the field at
infinity equals either to $h_{3}(\kappa)=\frac{3(10-\kappa)}{2\kappa}$
or to $h_{1}(\kappa)=\frac{6-\kappa}{2\kappa}$. This follows from CFT
fusion rules.  For reasons already explained we shall not mixed them.

The case $h_\infty=h_{3}(\kappa)$ is the simplest.
There is only one possible partition function with this scaling,
namely
$$  [(X_2-X_1)(X_3-X_1)(X_3-X_2)]^{2/\kappa}$$
It is expected to correspond to configurations with three curves starting
at initial positions $X_1,\ X_2$ and $X_3$ and converging toward infinity. 

\begin{figure}
\includegraphics[width=1.0\textwidth]{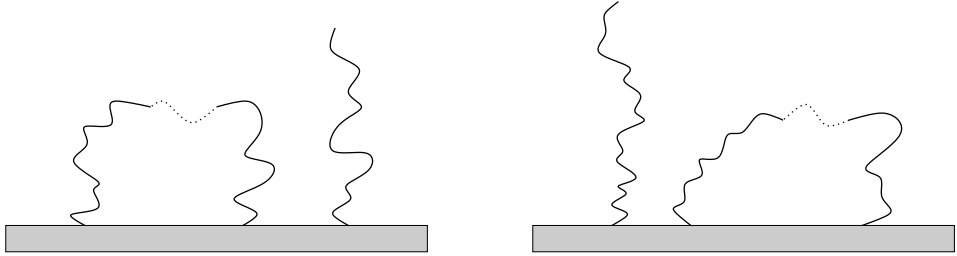}
\caption{\emph{For $h_\infty=h_1(\kappa)$ the curves of 3SLE join either
    $[X_1 X_2]$ and $[X_3 X_4]$ (on the left) or $[X_4 X_1]$ and
    $[X_2 X_3]$ (on the right).}}
\label{fig: 3SLEs}
\end{figure}

\begin{figure}[b]
\includegraphics[width=1.0\textwidth]{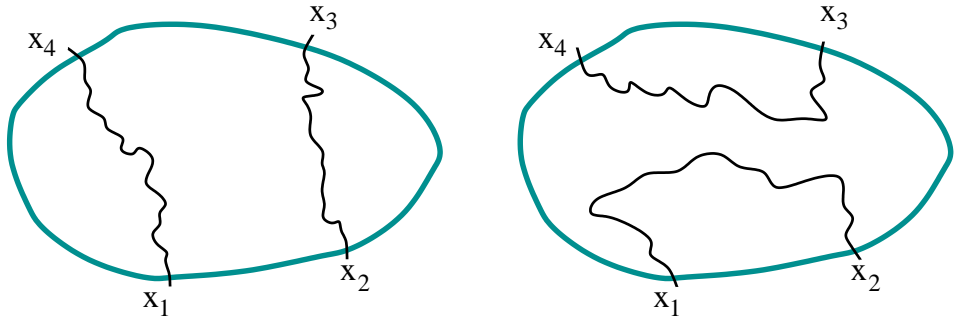}
\caption{\emph{Arch configurations for four SLE processes in 
an arbitrary domain.}}
\label{fig: 4SLEs}
\end{figure}

The case $h_\infty=h_{1}(\kappa)$ is more interesting since the space of
such partition functions is of dimension two and coincides with the
space of conformal block with 4 insertions of  boundary operators
$\psi$, with one localized at $X_4=\infty$:
$$
\langle \psi(X_4)\psi(X_3)\psi(X_2)\psi(X_1)\rangle
$$
We assume the points to be ordered $X_1<X_2<X_3<X_4$.  The
associated process should describe a family of two curves joining any
pair of adjacent points without crossing. There are thus two possible
topologically distinct geometries: either the curves join the pairs
$[X_1X_2]$ and $[X_3X_4]$ or they join $[X_4X_1]$ and $[X_2X_3]$,
see figure \ref{fig: 3SLEs}.  As expected, the number of topologically distinct
configuration equals that of conformal blocks, namely two.  Notice
that the last process is the same as a 4SLE but with the speed $a_4$
vanishing, see figure \ref{fig: 4SLEs}.

By conformal invariance we may normalize the points so that $X_1=0$,
$X_2=x$, $X_3=1$ and $X_4=\infty$ with $0<x<1$. We have two distinct
topological configurations and we thus have to identify the two
corresponding pure partition functions. This is will be done by
specifying the way the partition functions behave when points are
fused together. By construction these partition functions may be
written as correlation functions
$$Z(x)=\langle\psi(\infty)\psi(1)\psi(x)\psi(0)\rangle$$
so that their behavior when  points are fused are governed by CFT
fusion rules. As a consequence, $Z(x)$ behave either as
$x^{\frac{\kappa-6}{\kappa}}$ or as $x^{\frac{2}{\kappa}}$ as $x\to 0$.

We select the pure partition functions $Z_I$ and $Z_{II}$ by demanding that:
\begin{eqnarray}
Z_I(x) &=& x^{\frac{\kappa-6}{\kappa}}\times [1+\cdots],\quad 
~~~~~~~~~~~~ {\rm as}\ x\to0 
\label{bdrypure}\\
  &=& (1-x)^{\frac{2}{\kappa}}\times [{\rm const.}+\cdots],\quad 
~ {\rm as}\ x\to 1 \nonumber
\end{eqnarray}
and $Z_{II}(x)=Z_I(1-x)$ so that
\begin{eqnarray*}
Z_{II}(x) &=& x^{\frac{2}{\kappa}}\times [{\rm const.}+\cdots],\quad 
~~~~~~~~~~ {\rm as}\ x\to 0 \\
  &=& (1-x)^{\frac{\kappa-6}{\kappa}}\times [1+\cdots],\quad 
~~~~~~ {\rm as}\ x\to 1
\end{eqnarray*}
$Z_I$ will turn out to be the pure partition function for configurations in
which the curves join the pairs $[0x]$ and $[1\infty]$ while $Z_{II}$
will turn out to correspond to the configurations $[x1]$ and $[\infty 0]$.
The rationale behind these conditions consists in imposing that the
pure partition function possesses the leading singularity, with
exponent $(6-\kappa)/\kappa$, when $x$ is approaching the point allowed
by the configuration but has subleading singularity, with exponent
$2/\kappa$, when $x$ is approaching the point forbidden by the
configuration.

This set of conditions uniquely determines the functions $Z_I$ and
$Z_{II}$. These follows from CFT rules but may also be checked by
explicitly solving the differential equation that these functions
satisfy. Writing $Z(x)=x^{2/\kappa}(1-x)^{2/\kappa}\; G(x)$ yields,
$$ \kappa^2 x(1-x) G''(x) + 8\kappa (1-2x) G'(x) -
4(12-\kappa) G(x)=0 $$
so that $G(x)$ is an hypergeometric function and
$$
Z_{II}(x) = {\rm const.} x^{2/\kappa}(1-x)^{2/\kappa}\;
F(\frac{4}{\kappa}, \frac{12-\kappa}{\kappa};\frac{8}{\kappa}|x)
$$
with the constant chosen to normalize $Z_I$ as above.
Using this explicit formula one may verify that $Z_I(x)$ is
effectively a positive number for any $x\in[0;1]$ so it has all
expected properties to be a pure partition function.

For $\kappa=4$, $Z_I(x)=\sqrt{(1-x)/x}$
and for $\kappa=2$, $Z_I(x)=(1-x^2)/x^2$.

\subsubsection{Arch probabilities}
Let us now compute the probabilities for having one of the two
topologically distinct configurations: either $(I)$ with curves
joining either $[0x]$ and $[1\infty]$ or $(II)$ with curves joining
$[x1]$ and $[\infty0]$ as we just discussed. We shall proceed blindly,
but the reader should beware that there are subtleties involved.  What
is computed is the probability for certain $X^{(i)}_t$'s to hit each
other. What happens at the level of hulls and how the process should
be properly continued is not investigated, but is expected to yield
the announced probability for arch configuration.

We consider a generic partition function $Z$ which is the sum of the
pure partition functions $Z_I$ and $Z_{II}$:
$$ 
Z(x)= p_I Z_I(x) + p_{II}Z_{II}(x)
$$
with $p_I$ and $p_{II}$ positive. To specify the 3SLE (or 4SLE)
process we need the partition function $Z(X_1,X_2,X_3,X_4)$ which is
recover from $Z(x)$ by conformal transformation~:
$$
Z(X_1,X_2,X_3,X_4)= [(X_4-X_2)(X_3-X_1)]^\frac{\kappa-6}{\kappa}\;Z(X)
$$
with $X$ the harmonic ratio of the four points $X_1$, $X_2$, $X_3$ and $X_4$~:
$$X=\Big(\frac{ X_1-X_2 }{ X_1-X_3 }\Big)\Big(
\frac{ X_4-X_3 }{ X_4-X_2 }\Big).$$ 

Let $M_I(x)$ and $M_{II}(x)=1-M_I(x)$ be defined by
$$
 M_I(x) \equiv p_I Z_I(x)/Z(x)\quad,\quad
M_{II}(x)\equiv p_{II} Z_{II}(x) / Z(x) 
$$
By construction the processes $t\to M_I(X_t)$ and $t\to M_{II}(X_t)$,
with $X_t$ the harmonic ratio of the four moving points, are local
martingales. Since both $Z_I$ and $Z_{II}$ are positive, $M_I$ are
$M_{II}$ are bounded local martingales and thus are martingales.

Let $\tau$ be the stopping time given by the first instant at which a
pair of points $X_t^{(i)}$ coincide. Then, in configuration $(I)$ we
have $\lim_{t\nearrow \tau} X_t=0$ while $\lim_{t\nearrow \tau} X_t
=1$ in configuration $(II)$. Since, for $\kappa<8$, $M_I(x)$ is such
that $\lim_{x\to 0} M_I(x)= 1$ but $\lim_{x\to 1}M_I(x)=0$, we obtain
that $M_I$ evaluated at the stopping time $\tau$ is the characteristic
function for events with the topological configuration $(I)$, ie:
\begin{eqnarray*}
\lim_{t\nearrow \tau} M_I(X_t) &=& {\bf 1}_{{\rm config.} (I)} \\
\lim_{t\nearrow \tau} M_{II}(X_t) &=& {\bf 1}_{{\rm config.} (II)}
\end{eqnarray*} 
Since $M_I$ and $M_{II}$ are martingales, we get the probability of
occurrence of configurations of topological type $(I)$:
\begin{eqnarray}
\prob [{\rm config.} (I)]=M_I(X_{t=0})= \frac{p_I
  Z_I(x)}{p_IZ_I(x)+p_{II}Z_{II}(x)} 
\label{crossprob}
\end{eqnarray}
and similarly for the probabilities of having configuration $(II)$.
As expected they are ratios of partition functions.

\subsection{Applications to percolation and Ising model}
\label{sec: Ising}
We are now ready to give an application of triple (or
quadruple) SLE to percolation and Ising model.
Exploration processes in critical percolation are
described by SLEs with $\kappa=6$, as proved in
\cite{SS: percolation}.
Interfaces of spin clusters in critical Ising model is believed to
correspond to $\kappa = {3}$ while interfaces of Fortuin-Kasteleyn
clusters -- which occur in a high temperature expansion of the Ising
partition function -- are expected to correspond to the dual value
$\kappa=16/3$. 

What we have in mind are these statistical models, defined on the
upper half plane, with boundary condition changing operators at the four
points $0,\ x,\ 1$ and $\infty$. They change the boundary condition
from open to closed (or vice versa) in percolation $(\kappa=6)$ and
from plus to minus (or vice versa) for Ising model $(\kappa=3)$.

To apply previous results on 4SLE processes to these situations, we
have to specify the partition functions $Z(x)$, or equivalently, we
have to specify the value of $p_I$ and $p_{II}$. This is done by
noticing that these models are left-right symmetric so that for $x=1/2$
there is equal probability to find configuration $(I)$ or $(II)$.
Since $Z_I(1/2)=Z_{II}(1/2)$, we have $p_I=p_{II}=1$, so that the
total partition function is $Z(x)=Z_I(x)+Z_{II}(x)$ and the
probability of occurrence of configuration $(I)$ for any $0<x<1$ is
now:
$$
\prob [{\rm config.} (I)]= \frac{
  Z_I(x)}{Z_I(x)+Z_{II}(x)}\quad,\quad Z_{II}(x)=Z_I(1-x)
$$ 

\vspace{.2cm}

-- Percolation corresponds to $\kappa=6$. The boundary changing operator
$\psi$ has dimension $0$. The pure partition function $Z_I$ has a
simple integral representation:
$$
Z_I(x)_{\rm perco}= \frac{\Gamma(2/3)}{\Gamma(1/3)^2}\ \int_x^1 \ud
s\ s^{-2/3}(1-s)^{-2/3}.
$$
 By construction $Z_{II}(x)=Z_I(1-x)$
also possesses a simple integral representation but, most importantly,
it is such that the total partition function is constant,
$Z(x)=Z_I(x)+Z_{II}(x)=1$, as expected for percolation. As a
consequence we find:
$$
\prob [{\rm config.} (I)]_{\rm perco}=  
\frac{\Gamma(2/3)}{\Gamma(1/3)^2}\ \int_x^1 \ud s\ s^{-2/3}(1-s)^{-2/3}
$$
This is nothing but Cardy percolation crossing formula.

\vspace{.2cm}

-- Ising spin clusters correspond to $\kappa=3$. The boundary changing
operator $\psi$ has dimension $1/2$ and may thus be identified with a
fermion on the boundary. However the pure partition functions do not
correspond to the free fermion conformal block. By solving the
differential equation with the appropriate boundary
condition we get:
$$
Z_I(x)_{\rm spin\ Ising}= \mathrm{const}
\frac{1-x+x^2}{x(1-x)}\int_x^1\ud y\frac{(y(1-y))^{2/3}}{(1-y+y^2)^2}
$$
The total partition function $Z_I(x)+Z_I(1-x)$ is proportional to
$\frac{1-x+x^2}{x(1-x)}$, which is the free fermion result.

Hence, the Ising configuration probabilities are~:
$$
\prob [{\rm config.} (I)]_{\rm spin\ Ising}= \int_x^1\ud
y\frac{(y(1-y))^{2/3}}{(1-y+y^2)^2}\; \Big/ \int_0^1\ud
y\frac{(y(1-y))^{2/3}}{(1-y+y^2)^2}
$$
This is nothing but a new -- and previously unknown -- Ising crossing
formula. 

\vspace{.2cm}

-- FK Ising clusters correspond to $\kappa=16/3$. The operator $\psi$ has
then dimension $1/16$. The pure partition function are given by:
$$
Z_I(x)_{\rm FK\ Ising} = \frac{(1-x)^{3/8}}{x^{1/8}(1+\sqrt{x})^{1/2}}
$$
and the crossing probabilities by:
\begin{eqnarray*}
\prob [{\rm config.} (I)]_{\rm FK\ Ising} =
    \frac{\sqrt{(1-x) + (1-x)^{3/2}}}{\sqrt{x + x^{3/2}}
    + \sqrt{(1-x) + (1-x)^{3/2}}}
\end{eqnarray*}

The other critical random cluster (or Potts) models with $0 \leq Q
\leq 4$ have $Q = 4 \cos^2 \big( \frac{4 \pi}{\kappa} \big)$,
$4 \leq \kappa \leq 8$ and it is straightforward to obtain
explicit crossing formulas involving only hypergeometric functions.

\subsection{nSLEs and beyond}

We now comment on how to compute multiple arch probabilities for
general nSLEs.  This section only aims at giving some hints on how to
generalize previous computations.  So it shall be sketchy. It is clear
that the key point is to identify the pure partition functions -- once
this is done the rest is routine.  As exemplified above by
eq.(\ref{bdrypure}) this is linked to CFT fusions. The rules there were
that, for a given arch system, fusing two points linked by an arch
produces the dominant singularity which means that the two boundary
operators are fused on the identity operator, whereas fusing two
points not linked by an arch produces the subleading singularity which
means the fusion of the two boundary fields on the identity should
vanish. In general there could be a whole hierarchy of arches with
arches in the interior of others, i.e. with a family of self-surrounding
arches, the next encircling the previous.
So we are lead to propose the following rules.\\
For a given arch configuration:\\
--- The most interior pair of adjacent pair of points, say
$X_i,X_{i+1}$ in a family of self-surrounding arches fused into the
identity operator, so that the pure partition function evaluated at
$X_i\simeq X_{i+1}$ should be proportional to
$(X_{i+1}-X_i)^\frac{\kappa-6}{\kappa}$ times the pure partition
function associated to the arch system with the interior arch
$[X_iX_{i+1}]$ removed. Symbolically~:
$$
Z_{\rm pure}(\cdots, X_i\simeq X_{i+1},\cdots)\simeq  {\rm const.}\
(X_{i+1}-X_i)^\frac{\kappa-6}{\kappa}\times
Z_{{\rm pure}\setminus[X_iX_{i+1}]}(\cdots,\cdots)
$$
for $X_i$ and $X_{i+1}$ linked by an arch.\\
\noindent
--- The fusion on the identity of any pair of adjacent points 
not linked by an arch should vanish, so that the fusion of this
pair of points produces the subleading singularity. Symbolically~:
$$
Z_{\rm pure}(\cdots, X_i\simeq X_{i+1},\cdots)\simeq {\rm const.}\
(X_{i+1}-X_i)^\frac{2}{\kappa} +\cdots
$$
for $X_i$ and $X_{i+1}$ for not linked by an arch.

\begin{figure}
\includegraphics[width=1.0\textwidth]{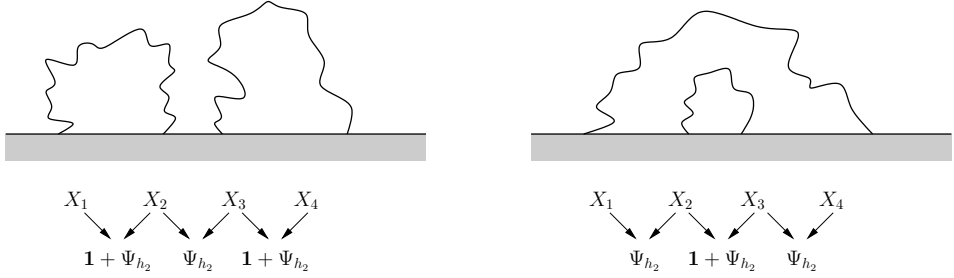}
\caption{\emph{Illustration of the fusion rules corresponding to
arch configurations.}}
\label{fig:rules}
\end{figure}

We do not have a complete proof that these rules fully determine the
pure partition functions but we checked it on a few cases, see figure
\ref{fig:rules}.

Here are a few samples. We shall give the relation between the pure
partition and the CFT conformal blocks indexed by the corresponding
Bratelli diagram.  For $n=4$, we may have the following arch systems
$[X_1X_2][X_3X_4]$ or $[X_1[X_2X_3]X_4]$.  (A given geometrical
configuration may correspond to different arch systems depending at
which location we open the closed boundary. But they are all
equivalent to these two up to an order preserving relabeling of the
points. For instance $[X_4X_1][X_2X_3]$ is equivalent to
$[X_1[X_2X_3]X_4]$.)  Applying the previous rules we get:
\begin{eqnarray*}
Z_{[X_1[X_2X_3]X_4]} &=& 
\langle {}_{[h_0{}]}\psi(X_1)
{}_{[h_1{}]}\psi(X_2)
{}_{[h_2{}]}\psi(X_3)
{}_{[h_1{}]}\psi(X_4)
{}_{[h_0{}]}\rangle
 \\
Z_{[X_1X_2][X_3X_4]}&=& 
\langle {}_{[h_0{}]}\psi(X_1)
{}_{[h_1{}]}\psi(X_2)
{}_{[h_0{}]}\psi(X_3)
{}_{[h_1{}]}\psi(X_4)
{}_{[h_0{}]}\rangle \\
 & +  & \omega \; 
\langle {}_{[h_0{}]}\psi(X_1)
{}_{[h_1{}]}\psi(X_2)
{}_{[h_2{}]}\psi(X_3)
{}_{[h_1{}]}\psi(X_4)
{}_{[h_0{}]}\rangle,
\end{eqnarray*}
where the indices $h_{m}$, $m=0,1,\cdots$ refer to the corresponding
points in the Bratelli diagram, i.e. to the weights $h_{m}(\kappa)$ of
the intermediate Virasoro modules. The coefficient $\omega$ is fully
determined, in terms of CFT fusion coefficients, by demanding that the
fusion of $X_2$ and $X_3$ on the identity vanishes.

One may go on and solve for the pure partition functions in few
other cases. A particularly simple example with $n=6$ is given by~:
\begin{eqnarray*}
Z_{[X_1[X_2[X_3X_4]X_5]X_6]} & = & \\
& & \hspace{-5cm}
\langle {}_{[h_0{}]}\psi(X_1)
{}_{[h_1{}]}\psi(X_2)
{}_{[h_2{}]}\psi(X_3)
{}_{[h_3{}]}\psi(X_4)
{}_{[h_2{}]}\psi(X_5)
{}_{[h_1{}]}\psi(X_6)
{}_{[h_0{}]}\rangle
 \end{eqnarray*}
As can be seen on these examples, there is no simple relation between
arch systems and Bratelli diagrams and the change of basis for one
to the other is quite involved. The only simple rule we find is that
the pure partition function for a unique family of self-surrounding
arches is a pure conformal block corresponding to a unique Bratelli
diagram.

\end{document}